\documentclass[12pt]{article}

\usepackage{xr}
\usepackage{natbib}
\usepackage{amsmath}
\usepackage{graphicx}
\usepackage{enumerate}
\usepackage{bm}
\usepackage{amssymb}
\usepackage[usenames,dvipsnames]{xcolor}
\usepackage{amsmath}
\usepackage[ruled]{algorithm2e}
\usepackage{caption}
\usepackage{url}
\usepackage[margin=1.2in]{geometry}

\usepackage{booktabs} %

\usepackage{setspace}
\begin{document}

\title{A multivariate Gaussian random field prior against spatial confounding}

\author{Isa Marques$^1$ \\ \small{imarques@uni-goettingen.de}
   \and   Thomas Kneib$^1$ \\ \small{tkneib@uni-goettingen.de}
   \and   Nadja Klein$^2$ \\  \small{nadja.klein@hu-berlin.de} }

\date{
\small
$^1$ Georg-August-University of G\"{o}ttingen, Chairs of Statistics and Econometrics, Humboldtallee 3, 37073 G\"{o}ttingen, Germany \\
$^2$ Humboldt-Universit\"at zu Berlin, School of Business and Economics, Unter den Linden 6, 10099 Berlin, Germany
}

\maketitle

\begin{abstract}
Spatial models are used in a variety research areas, such as environmental sciences, epidemiology, or physics.
A common phenomenon in many spatial regression models is spatial confounding. This phenomenon takes place when spatially indexed covariates modeling the mean of the response are correlated with the spatial random effect. As a result, estimates for regression coefficients of the covariates can be severely biased and interpretation of these is no longer valid.
Recent literature has shown that typical solutions for reducing spatial confounding can lead to misleading and counterintuitive results.
In this paper, we develop a computationally efficient spatial model in a Bayesian framework integrating novel prior structure that reduces spatial confounding.
Starting from the univariate case, we extend our prior structure to case of multiple spatially confounded covariates.
 In a simulation study, we show that our novel model flexibly detects and reduces spatial confounding in spatial datasets, and it performs better than typically used methods such as restricted spatial regression. These results are promising for any applied researcher who wishes to interpret covariate effects in spatial regression models. As a real data illustration, we study the effect of elevation and temperature on the mean of daily precipitation in Germany.
\end{abstract}

\noindent%

\section{Introduction}

Consider a set of observations $\lbrace y(\bm{s}): \bm{s} \in \mathcal{S} \rbrace$, where $\bm{s}$ is a spatial index variable within a spatial domain $\mathcal{S}$. For spatial data, observations tend to exhibit dependence simply because they are close to each other. In a regression model context, this spatial dependence is typically taken into account by including an additional error term, $\gamma(\cdot)$, in the regression equation such that
\begin{equation}\label{eq:model_equation}
  y(\bm{s}_i) = \beta_0 + \bm{z
  }(\bm{s}_i)^T\bm{\beta} + \gamma(\bm{s}_i) + \varepsilon_i, \quad i = 1, \ldots, n,
\end{equation}
where $n$ is the number of observations. The vector $\bm{z}(\bm{s}_i)$ consists of $B$ spatially indexed covariates evaluated at location $\bm{s}_i \in \mathcal{S}$, with associated coefficient vector $\bm{\beta} = ( \beta_1, ...,
\beta_B)^T$. Moreover,
$\varepsilon_i \sim N(0,\sigma_\varepsilon^2)$ is an \textit{i.i.d.}\ non-spatial error or nugget.

It is generally believed that one should account for spatial correlation when dealing with spatial data \citep{Cressie_1993}. Otherwise, statistical inference for regression coefficients might be incorrect and prediction errors high \citep{Cressie_1993, hoeting2006model}.
Consequently, spatial models have become common in many disciplines where data are typically spatial in nature, e.g., epidemiology, ecology, or geology.

While spatial models offer a very popular and flexible approach for modeling spatial data, they also suffer from two major shortcomings. Namely: (i) the computational challenges posed by high-dimensional latent variables, and (ii) the lack of interpretable regression coefficients $\bm{\beta}$ in \eqref{eq:model_equation} due to spatial confounding.

The first shortcoming relates to one of the main challenges in spatial statistics: how to handle large data sets. In this paper, we consider a continuous spatial domain $\mathcal{S}$.
The most common random object for representing continuously indexed spatial effects, $\gamma(\bm{s})$, are Gaussian random fields (GRFs). GRFs can conveniently be characterized by their mean function $\mu_\gamma(\bm{s})$ and covariance function $\Sigma_{\bm{\gamma}}(\bm{s}, \bm{s}')$, with $\bm{s}, \bm{s}' \in \mathcal{S}$. However, likelihood evaluation and spatial predictions are in general cubic in the number $n$ of observations of a GRF, making GRFs especially expensive for large datasets.

 The second shortcoming relates to the fact that in models such as \eqref{eq:model_equation}, covariates that vary spatially, $\bm{z}(\bm{s})$, are often correlated with each other, but also with the spatial random effect, $\gamma(\bm{s})$.
While the former phenomenon is typically denoted as multicollinearity, the latter is denoted as spatial confounding.
Spatial confounding was first identified by \citet{clayton1993spatial}, who referred to it as ``confounding by location'': the situation where estimates of a regression coefficient associated with a spatially-structured covariate are affected by the presence of a spatial random effect in the model.
More recently, \citet{hanks2015restricted} and \citet{nobre2020effects} argued that if both $z(\bm{s})$ and $\gamma(\bm{s})$ are spatially smooth, they are collinear even if they are stochastically independent.

While collinearity between covariates in the fixed effects of the model has been extensively discussed in the literature, collinearity between these covariates and spatial random effects is mentioned less frequently.
This is in part because the spatial random effect is unobservable, making it difficult to clearly separate its effect.
Nonetheless, as highlighted by \citet{reich2006effects} and \citet{hodges2010adding}, spatial confounding should be taken into account for, as it biases the regression coefficients and ``overinflates'' the uncertainty associated with these estimates.  In a Bayesian context, confounding reduces the sampler's efficiency \citep{paciorek2010importance}. In contrast, spatial confounding leaves predictions unchanged, or even improves them~\citep{page2017estimation}.

One typical solution for spatial confounding, initially brought forward by \citet{reich2006effects} for conditional autoregressive models and discrete $\mathcal{S}$, is denoted in this paper as restricted spatial regression (RSR).
RSR  alleviates confounding by projecting the spatial effects into the orthogonal space spanned by the covariates. This idea has been extended to continuous $\mathcal{S}$, e.g., by \citet{hanks2015restricted} and \citet{ingebrigtsen2014spatial}.
However, using orthogonality to avoid bias from accidentally accounting for some of the effect of the covariate in the residual might be too restrictive.
Recent studies show that the use of RSR has counterintuitive consequences which defy the general expectations in the literature. Namely, \citet{khan2020restricted} unfold that, for discrete $\mathcal{S}$,  RSR typically leads to poorer coverage rates than non-spatial methods, while
\citet{hanks2015restricted} point out that, in case of for continuous $\mathcal{S}$, RSR offers increased computational efficiency at the cost of significantly poorer coverage rates than a spatial model that does not account for spatial confounding.

Some alternatives to RSR have been proposed in the literature.
For discrete $\mathcal{S}$, \citet{thaden2018structural} developed a structural equation model that reduces confounding by including an additional structural equation that models the spatial structure of $z(\bm{s})$.
For continuous $\mathcal{S}$, \citet{chiou2019adjusted} assume that a confounded covariate $z(\bm{s})$ follows a Gaussian distribution and model its correlation with $\gamma(\bm{s})$.
By adjusting the RSR estimator for this correlation, the bias of regression coefficients is reduced. However, the authors claim there is no way of estimating the correlation parameter within the model in a frequentist framework. As an alternative, they provide an heuristic discussion regarding the performance of their method based on a simulation study where the correlation parameter is unknown. Neither \citet{thaden2018structural} or \citet{chiou2019adjusted} provide an in-depth discussion of the procedure when multiple spatially indexed covariates are correlated with the spatial random effect.

Finally, some authors have developed models that simultaneously alleviate the computational bottlenecks of dealing with high dimensional latent effects and spatial confounding. It is within this group that we wish to develop our work.
\citet{hughes2013dimension} present a reduced-rank approach for areal data that uses the eigenvectors of the Moran I operator \citep{moran1950notes} as synthetic predictors.
Another computationally efficient approach to implement RSR for continuous data without dimensionality reduction within a Markov chain Monte Carlo (MCMC) sampler is to constrain $\gamma(\bm{s})$ to be orthogonal to the fixed effects by ``conditioning by Kriging'' \citep{rue2005gaussian}. For further computational benefits, this can be combined with the computationally efficient stochastic partial differential equation (SPDE) approach \citep{lindgren2011explicit}.

 This paper develops a Bayesian framework that works against spatial confounding in continuously indexed spatial models using the SPDE-approach and a novel prior structure.
The issue of spatial confounding in models such as \eqref{eq:model_equation} is also, intrinsically, an identification problem between fixed and spatial random effects \citep{paciorek2010importance}.
This identification problem can be improved through constraints on $\gamma(\bm{s})$, for example, through the prior on $\gamma(\bm{s})$ \citep{paciorek2010importance}. By imposing structure through a prior distribution in a non-identified model we can help accounting for a portion of the confounding, thus potentially improving the bias of the estimators. In this paper, we explicitly account for the correlation between the spatial random effect $\gamma(\bm{s})$ and - potentially multiple - covariates $\bm{z}(\bm{s})$ by defining a new prior structure for $\gamma(\bm{s})$, which we call a \emph{multivariate Gaussian random field prior}. Moreover, we explore a penalized complexity (PC) prior for the correlation parameter that controls the shrinkage towards  a base model, i.e.~the case of no spatial confounding \citep{simpson2017penalising}.
In summary, with this paper, we
\begin{enumerate}
  \item develop a new prior structure against spatial confounding,
  \item increase computational efficiency of the developed prior by exploiting the sparsity of the precision matrix of GMRFs in the SPDE-approach, and
  \item  extend the prior structure to the case of multiple covariates being correlated with the spatial random effect.
\end{enumerate}

The remainder of this paper is organized as follows. In Section~\ref{sec:method}, we introduce the novel computationally efficient prior structure against spatial confounding. Section~\ref{sec:posterior} elaborates on the posterior estimation of the full hierarchical model. In Section~\ref{sec:simulation}, we present an in-depth simulation study that investigates the sources and consequences of spatial confounding. This simulation study compares our model with the base model, RSR and a non-spatial model. In Section~\ref{sec:application}, we demonstrate the useful properties of our prior structure in an application to precipitation data from Germany.
Section~\ref{sec:discussion} concludes.

\section{Model structure}\label{sec:method}
In this section, we present a novel prior structure for reducing spatial confounding in model \eqref{eq:model_equation}, although the concept can be transferred over to other response families. %
In what follows, we first introduce the SPDE-approach, our  method of choice for estimating GRFs due to its computational benefits in Section~\ref{sec:spde}. In Section~\ref{sec:newprior}, we then present our novel prior structure for the case of only one covariate correlated with the spatial random effect. Finally, Section~\ref{sec:multivariate} extends this prior to the case of multiple covariates correlated with the spatial random effect.

\subsection{The SPDE approach to Gaussian random fields}\label{sec:spde}
A spatial process  $\lbrace \gamma(\bm{s}): \bm{s} \in \mathcal{S} \rbrace$ is a continuously indexed GRF if all finite-dimensional distributions of the process are Gaussian, i.e.,
for all $n \in \mathbb{N}$ and all $\bm{s}_1, \dots, \bm{s}_n \in \mathcal{S}$, the vector $(\gamma(\bm{s}_1), \dots, \gamma(\bm{s}_n))^{T}$ is multivariate Gaussian distributed. A GRF is specified by using a mean function $\mu_{\bm{\gamma}}(\bm{s})$ and a covariance function $\Sigma_{\bm{\gamma}}(\bm{s}, \bm{s}')$, for $\bm{s}, \bm{s}' \in \mathcal{S}$. Throughout this paper, we assume $\mu_{\bm{\gamma}}(\bm{s}) = 0$ for all $\bm{s}$.

 Gaussian Markov random fields (GMRFs) are essentially multivariate Gaussian distributions satisfying conditional independence assumptions, i.e., they have a sparse precision matrix, $\bm{Q}_{\bm{\gamma}} = \bm{\Sigma}_{\bm{\gamma}}^{-1}$. While GMRFs are sometimes less attractive than GRFs from a theoretical viewpoint, they can make up for it by decreasing computational costs.
 \citet{rue2002fitting} demonstrated empirically that GMRFs can closely approximate most of the commonly used covariance functions in geostatistics and proposed using them as computational replacements for GRFs for computational reasons. This method is denoted SPDE-approach \citep{lindgren2011explicit}.

The link between the GRF and GMRFs in Mat\'{e}rn models is achieved via the SPDE that gives name to the method.  Namely,
\begin{equation*}%
(\kappa_{\bm{\gamma}}^2 - \Delta)^{{\alpha}_{\bm{\gamma}}/2} ( \tau_{\bm{\gamma}}\, \gamma(\bm{s})) =
\mathcal{W}(\bm{s}), \quad \bm{s} \in \mathcal{S}, \quad \alpha_{\bm{\gamma}}=
\nu_{\bm{\gamma}} +  1,
\end{equation*}
where $\Delta$ is the Laplacian operator and $\mathcal{W}$ is a Gaussian spatial white
noise innovation process. Moreover,  $\tau_{\bm{\gamma}} > 0$ is a precision parameter, $\kappa_{\bm{\gamma}} > 0$ is related to the spatial range and $\nu_{\bm{\gamma}} > 0$ is a smoothness parameter. Throughout this paper, we assume $\mathcal{S} \subseteq \mathbb{R}^2$.

The obtain the solution
$\gamma(\bm{s})$ of the SPDE on $\mathcal{S}$, which is a stationary
GRF with Mat\'{e}rn covariance function,
 we first discretize the domain using the finite elements method (FEM), i.e., a triangulation of the domain.
Specifically, the GRF is
expanded into a piecewise linear basis through
\begin{equation}\label{eq:linearbasisfunctions}
\gamma(\bm{s}) = \sum^M_{m=1} \psi_m(\bm{s}) \gamma_m,
\end{equation}
where the joint distribution of the weight vector $\bm{\gamma} = (\gamma_1,
\ldots, \gamma_M)'$ is chosen so that the distribution of the functions
$\gamma(\bm{s}) $ approximates the distribution of solutions to the SPDE on the
domain.
Each basis function $\psi_m(\cdot)$ takes the value of one on node $m$ and the value zero on the nodes of
the neighboring triangles of the triangulated domain, being piecewise linear on these triangles.

The FEM solution leads to normally distributed weights
$\bm{\gamma}$ with mean zero and sparse precision matrix $\bm{Q}_{\bm{\gamma}}$ such
that $\bm{\gamma} \sim N(\bm{0}, \bm{Q}_{\bm{\gamma}}^{-1}).$ That is, $\bm{\gamma}$ is a GMRF that approximates the stationary GRF that solves the SPDE. A sparse precision matrix $\bm{Q}_{\bm{\gamma}}$ will typically reduce the cost of factorizations - such as the Cholesky decomposition - to $O(n^{3/2})$ in spatial models, instead of $O(n^{3})$ for GRFs with dense covariance matrices. Moreover, the Cholesky decomposition of the precision $\bm{Q}_{\bm{\gamma}}$ can inherit its sparse structure. Namely, if the mesh nodes are ordered such that $\bm{Q}_{\bm{\gamma}}$ is a band matrix with bandwidth $p$, then the Cholesky triangle of $\bm{Q}_{\bm{\gamma}}$ has (lower) bandwidth $p$ \citep{rue2005gaussian}. This property can be exploited in computational efficient methods.

The marginal variance $\sigma_{\bm{\gamma}}^2$  of the GRF can be derived from the SPDE's parameters $\tau_{\bm{\gamma}}$ and $\kappa_{\bm{\gamma}}$, via
\begin{equation*}%
\sigma_{\bm{\gamma}}^2 = \frac{\Gamma(\nu_{\bm{\gamma}})}{\Gamma(\alpha_{\bm{\gamma}}) 4 \pi \kappa_{\bm{\gamma}}^{2 \nu_{\bm{\gamma}}} \tau_{\bm{\gamma}}^2}.
\end{equation*}
Furthermore, by defining the range as the distance at which the spatial correlation falls to
0.05 \citep{lindgren2011explicit}, the range parameter $r_\gamma$ can be written as
\begin{equation*}%
r_{\bm{\gamma}} = \frac{\sqrt{8 \nu_{\bm{\gamma}}}}{\kappa_{\bm{\gamma}}}.
\end{equation*}
We refrain from going into further detail on the specificities of the SPDE-approach in this paper, but the interested reader can find a in-depth account in \citet{lindgren2011explicit}, as well as an introduction to GMRFs in \citet{rue2005gaussian}.

\subsection{A novel MGRF prior: the one covariate case}\label{sec:newprior}
Throughout Section~\ref{sec:newprior}, we omit the spatial index $\bm{s}$. This is related with the notation used for GMRFs in \eqref{eq:linearbasisfunctions} and the reason of its use should become clear in Section~\ref{sec:re_1}.

Consider the model from \eqref{eq:model_equation}.
Let $\bm{z}$ be a covariate having a spatial structure that can be represented by a GRF with mean $\bm{\mu_Z}$ and covariance matrix $\bm{\Sigma_Z}$. That is,
$ \bm{z} \sim \mathcal{N}(\bm{\mu_z}, \bm{\Sigma_z}).$
Let $\bm{\gamma} \sim \mathcal{N}(\bm{0}, \bm{\Sigma_\gamma})$. Then, $\bm{\gamma}$ and $\bm{z}$ are jointly Gaussian with the following structure
\begin{align*}
\begin{pmatrix}
\bm{\gamma}\\
\bm{z}
\end{pmatrix} &\sim  \mathcal{N}
\begin{bmatrix}
\begin{pmatrix}
\bm{0}\\
\bm{\mu_z}
\end{pmatrix}\!\!&,&
\begin{pmatrix}
\bm{\Sigma_\gamma} & \rho \bm{\Sigma}_{\bm{\gamma}}^{1/2} (\bm{\Sigma}_{\bm{z}}^{1/2})^T \\
 \rho \bm{\Sigma_z}^{1/2} (\bm{\Sigma_\gamma}^{1/2})^T & \bm{\Sigma_z}
\end{pmatrix}
\end{bmatrix},
\end{align*}
where
$\rho \bm{\Sigma}_{\bm{\gamma}}^{1/2} (\bm{\Sigma}_{\bm{z}}^{1/2})^T = \bm{\Sigma}_{\bm{\gamma}\, \bm{z}}$, $\bm{\Sigma_{\bm{z}\, \bm{\gamma}}} = \bm{\Sigma}_{\bm{\gamma}\, \bm{z}}^T$, and $\rho \in [-1,1]$.
The matrix roots $\bm{\Sigma}_{\bm{\gamma}}^{1/2}$ and $\bm{\Sigma}_{\bm{z}}^{1/2}$ satisfy $\bm{\Sigma}_{\bm{\gamma}} = \bm{\Sigma}_{\bm{\gamma}}^{1/2} (\bm{\Sigma}_{\bm{\gamma}}^{1/2})^T$ and $\bm{\Sigma}_{\bm{z}} = \bm{\Sigma}_{\bm{z}}^{1/2} (\bm{\Sigma}_{\bm{z}}^{1/2})^T$, and can be derived from a Cholesky decomposition or an eigenvalue decomposition. The covariance structure is constructed such that $\mbox{Corr}(\gamma_m, Z_m) = \rho$, for all $m = 1, \ldots, M$.

\begin{figure}[tb]
 \center
  \includegraphics[width = 0.3\textwidth]{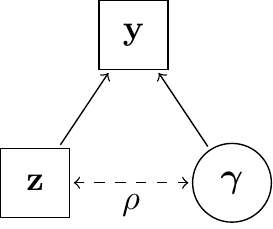}
  \caption{Visual representation of the prior structure. Squares represent observable variables and circles represent non-observables.}\label{fig:structural_Eq.}
 \end{figure}

The conditional distribution of $\bm{\gamma}$ given $\bm{z}$ is a Gaussian distribution with mean and covariance
\begin{align}\label{eq:mean_cond}
  \bm{\mu}_{\bm{\gamma} \vert \bm{z}} &= \bm{0} +  \bm{\Sigma}_{\bm{\gamma}\, \bm{z}}\bm{\Sigma}_{\bm{z}}^{-1}(\bm{z} - \bm{\mu}_{\bm{z}}),\\
\label{eq:covariance_cond}
    \bm{\Sigma}_{\bm{\gamma} \vert \bm{z}} &= \bm{\Sigma}_{\bm{\gamma}} - \bm{\Sigma}_{\bm{\gamma}\, \bm{z}}\bm{\Sigma}_{\bm{z}}^{-1} \bm{\Sigma}_{\bm{z}\, \bm{\gamma}}.
\end{align}
Putting it all together, our novel prior distribution follows
\begin{equation}\label{eq:cond_dist}
\bm{\gamma} \vert \bm{z} \sim \mathcal{N}(\bm{\mu}_{\bm{\gamma} \vert \bm{z}},  \bm{\Sigma}_{\bm{\gamma} \vert \bm{z}}) \quad \text{and} \quad \bm{z} \sim \mathcal{N}(\bm{\mu_z}, \bm{\Sigma_z}).
 \end{equation}
If $\rho = 0$, then $\bm{\gamma} \vert \bm{z} \sim \mathcal{N}(\bm{0}, \bm{\Sigma}_{\bm{\gamma}})$, i.e., we get the typically used  prior in continuously indexed spatial models, which does not account for confounding. We refer to a prior that assumes $\rho = 0$ as the base prior, and the model associated with it as the base model. The reason for this name should become clearer in Section~\ref{sec:rho}. The more general case with $\rho \in [-1,1]$ in \eqref{eq:cond_dist} is denoted a \emph{multivariate Gaussian random field (MGRF) prior}, and the model associated with it is a MGRF model.
The final MGRF prior structure is illustrated graphically in Figure \ref{fig:structural_Eq.}.

We assume $\bm{\Sigma}_{\bm{\gamma}}^{1/2}$ represents the Cholesky decomposition of $\bm{\Sigma}_{\bm{\gamma}}$. This allows us to exploit the fact that the Cholesky decomposition of the precision of the GRMF can inherit its sparse structure. This fact reduces considerably the computational complexity of the resulting prior structure, as explained in more detail in Sections \ref{sec:re_1} and \ref{sec:re_2}.

After some considerations on the mesh design in Section~\ref{sec:mesh}, we investigate reformulations of \eqref{eq:mean_cond}  in Sections \ref{sec:re_1} and \ref{sec:re_2} respectively. The former allows for considerable computational efficiency, while the latter is tailored to enhance  numerical stability. Finally, in Section \ref{sec:rho} we discuss our prior choice for $\rho$.

\subsubsection{Computational efficiency: mesh design}\label{sec:mesh}
Although the SPDE-approach does not require dimensionality reduction to accrue computational benefits,  the first obvious choice when trying to reduce the computational cost of computing the  prior is to reduce the number of mesh nodes $M$. Ultimately, for $M \ll n$ the proposed prior can be introduced at virtually no computational cost, but possibly with noticeable reduction in the accuracy of the approximation of the spatial random effect.

Therefore, we can devise solutions that represent a trade-off between computational efficiency and the approximation's accuracy. For example:
\begin{enumerate}
  \item Use $M \ll n$, achieving dimensionality reduction.
  \item Use a lower $M$, but still $M > n$.
\end{enumerate}
In Section~\ref{sec:simulation} and in Section~\ref{sec:application}, we use $M \approx n$.

\subsubsection{Computational efficiency: reformulation I}\label{sec:re_1}

In what follows, we will focus on $\bm{\gamma}$, but the same logic applies to $\bm{z}$.  For $\bm{\Sigma}_{\bm{\gamma}}$ positive definite, there is a unique Cholesky triangle $\tilde{\bm{L}}_{\bm{\gamma}}$ such that $\tilde{\bm{L}}_{\bm{\gamma}}$ is a lower triangular matrix with $\tilde{L}_{\bm{\gamma}(ii)} > 0\ \forall i$ and $\bm{\Sigma}_{\bm{\gamma}} = \tilde{\bm{L}}_{\bm{\gamma}}\ \tilde{\bm{L}}_{\bm{\gamma}}^T$. A GMRF is expressed in terms of  its sparse precision matrix $\bm{Q}_{\bm{\gamma}}$. Consequently, for GMRFs, it makes more sense to think in terms of the Cholesky decomposition of the precision matrix $\bm{Q}_{\bm{\gamma}} = \bm{L}_{\bm{\gamma}}\ \bm{L}_{\bm{\gamma}}^T$, i.e., $\bm{\Sigma}_{\bm{\gamma}} = (\bm{L}_{\bm{\gamma}}\ \bm{L}_{\bm{\gamma}}^T)^{-1} = \bm{L}_{\bm{\gamma}}^{-T} \bm{L}_{\bm{\gamma}}^{-1}$ where $ \bm{L}_{\bm{\gamma}}^{-T} = (\bm{L}_{\bm{\gamma}}^{-1})^{T}$.

The SPDE-approach uses a GMRF for computations. Consequently, throughout this paper we will always think in terms of precision matrices, rather than covariance matrices.
Omitting the spatial index $\bm{s}$ now gains some meaning, as we are referring to GMRFs and not GRFs (see Section \ref{sec:spde}). To go from the triangulated domain to the observation space we project $\bm{\gamma}$ into the correct space with the projection matrix $\bm{\psi}(\cdot)$ according to \eqref{eq:linearbasisfunctions}. Exploiting the Markov graph structure of the GMRF allows to divide its graph recursively into conditionally independent sets. This helps making the Cholesky factor of the precision matrix sparse. With this in mind, we can simplify \eqref{eq:mean_cond} and \eqref{eq:covariance_cond} and obtain
\begin{align}\label{eq:mean_cond_simplified}
  \bm{\mu_{\bm{\gamma} \vert \bm{z}}} &= \bm{0} + \rho\,                                                                                                                                                                                                                                                                      \bm{L}_{\bm{\gamma}}^{-T} \bm{L}_{\bm{z}}^{T}(\bm{z} - \bm{\mu_Z})\\
\bm{Q_{\bm{\gamma} \vert \bm{z}}} &= \bm{\Sigma_{\bm{\gamma} \vert \bm{z}}}^{-1} = \frac{1}{1 - \rho^2} \bm{Q_{\bm{\gamma}}}\label{eq:precision_cond_simplified}
\end{align}
as derived in detail in Supplement~\ref{app:ref_1}.
This reformulation of the prior's first and second order structure takes advantage of the sparse structure of $\bm{Q_\gamma}$ and $\bm{Q_z}$, and it can lead to a considerable decrease in the computational cost of factorizations. To guarantee positive definiteness, we need to avoid boundary cases $| \rho | \approx 1$. For values of $\rho$ close to one in absolute value, reformulation II should be used.

\subsubsection{Computational efficiency: reformulation II}\label{sec:re_2}

Building on reformulation I, one may alternatively consider the unconstrained GMRF, $\bm{\gamma} \sim \mathcal{N}(\bm{0}, \bm{Q_\gamma}^{-1})$, and then compute
\begin{equation*}
 \bm{\gamma}^* =  \bm{\gamma} + \rho\, \bm{L}_{\bm{\gamma}}^{-T} \bm{L}_{\bm{z}}^{T}( \bm{z} -   \bm{\gamma_z})
\mbox{,
where }
  \bm{\gamma_z} \sim \mathcal{N} (\bm{\mu_z}, \bm{\Sigma_z}
  ).
\end{equation*}
The resulting $\bm{\gamma}^*$ has the correct conditional distribution (see Supplement~\ref{app:ref_2}).
This procedure is similar to ``conditioning by Kriging'' in \citet{rue2005gaussian}.

Comparatively to reformulation I, this reformulation requires the sampling of one additional parameter $\bm{z}$ of size $M$. This can increase considerably the computational costs associated with MCMC (see Section~\ref{sec:sampler}), as $\bm{z}$ is sampled every iteration. However, when $| \rho | \approx 1$, only reformulation II can be used, as the precision matrix in reformulation I is no longer positive definite, and thus it is invalid. As we typically do not have information about the values $\rho$ takes, reformulation II is generally recommended.

\subsubsection{Prior for $\rho$}\label{sec:rho}
Estimating the correlation between $\bm{z}$ and $\bm{\gamma}$ is a complex task, namely because $\bm{\gamma}$ is an unobservable random effect.
In this paper, we estimate $\rho$ within the Bayesian hierarchical model presented in more detail in Section \ref{sec:posterior}, and set a PC-prior for $\rho$. In this way, we cover the gap in the work by \citet{chiou2019adjusted}, where the authors claim that in a frequentist setting there is no method to estimate $\rho$ within the model.

The concept of a PC-prior was first developed by \citet{simpson2017penalising} and an in-depth introduction can be found there. In short, as the name penalized complexity indicates, this prior invokes the principle of parsimony, for which a simpler \textit{base} model should be preferred until there is enough support for a more complex model. The prior's density decays at a constant decay-rate as a function of a measure of the increased complexity between the more flexible model and the base model.
The resulting prior allows for user-defined scaling: an upper bound for the parameter of interest can be defined, which describes what is considered as “tail event” and what is the weight put on this event. Hence, the user can describe how informative the resulting PC-prior is.

\citet{simpson2017penalising} developed a PC-prior for correlation parameters. Here, we consider a base model with  $\rho_0 = 0$, i.e., a base model that assumes no spatial confounding between fixed effects and spatial random effects. Let ${\frac{-1}{w -1} < \rho < 1}$ with  $w \geq 2$. \citet{simpson2017penalising} show that the PC-prior for $\rho$ has density
$$
p(\rho) = \frac{w - 1}{2} \left( \frac{1}{1 - \rho} - \frac{1}{1 + (w - 1)\rho } \right) \frac{\lambda}{\sqrt{- lR(\rho)}} \exp \left(-\lambda \sqrt{ - lR(\rho)} \right),
$$
where $lR(\rho) = \log(R(\rho))$ and $R(\rho) = ( 1 + (w - 1)\rho)(1 - \rho)^{w - 1}$.

\begin{figure}[ht!]
  \includegraphics[width=\textwidth]{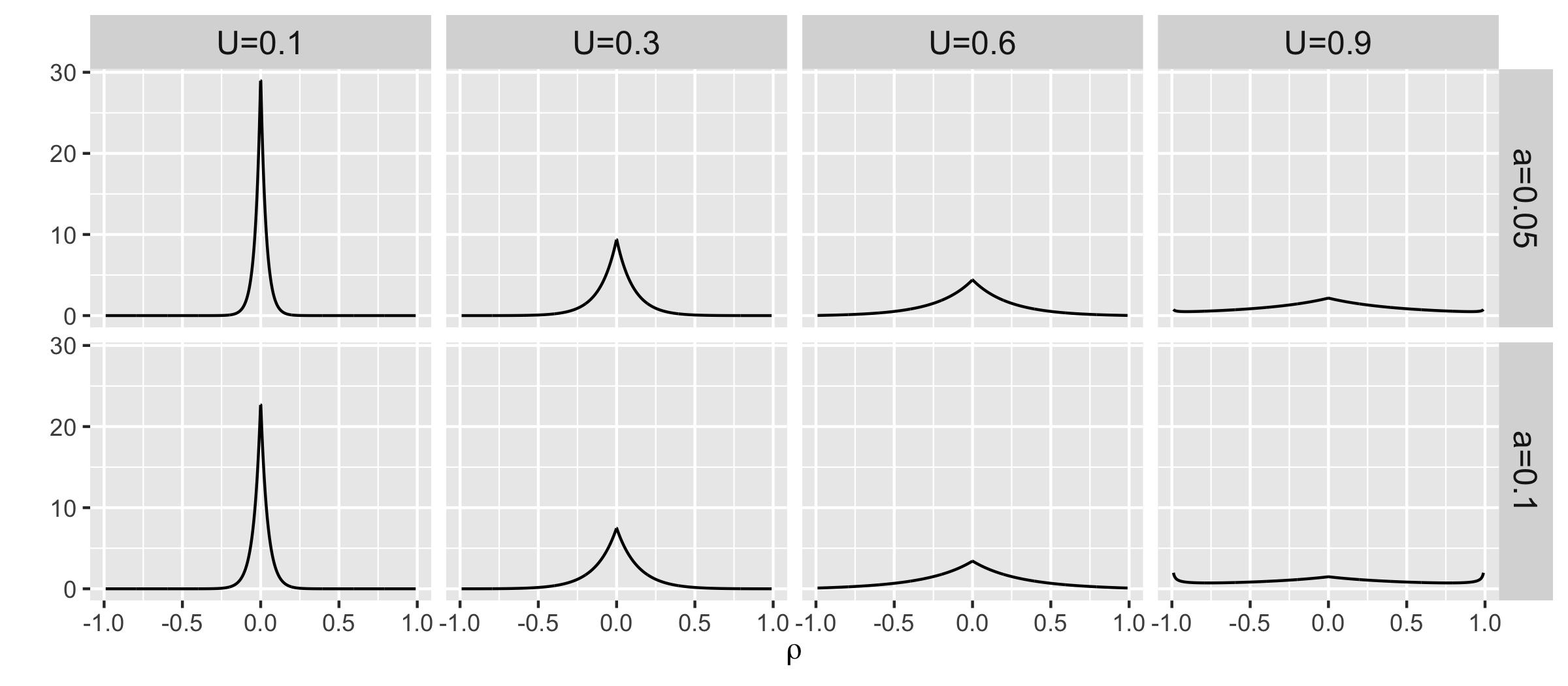}
  \caption{Prior densities for $\rho$ using a PC-prior with $\rho_0 = 0$, $w=2$ and different values for $U$ and $a$.}\label{fig:pc_rho}
\end{figure}

The decay-rate $\lambda$ can be chosen by sampling PC-priors from various values of $\lambda$ and choosing a $\lambda$ satisfying
\begin{equation}\label{eq:lambda}
  \mbox{Prob}(\vert \rho \vert > U ) = a.
\end{equation}
The resulting PC-prior is shown in Figure~\ref{fig:pc_rho}, for $w = 2$ and different values of $U$ and $a$. As it can be observed, the prior can be made less informative by increasing $U$ or $a$, i.e., by reducing shrinkage towards the base model. For different values of $w$, the left boundary for the correlation can be shifted towards zero, resulting in an asymmetric prior distribution for $\rho$. However, throughout this paper, we assume there is no information concerning the type of correlation between the unobservable spatial random field and the covariate. Thus, we only consider the case $w=2$. Moreover, herein, we set $a = 0.05$ and simply change $U$ to get the desired prior density. We refer to the resulting PC-prior as $\mathcal{PC}(U)$

\subsubsection{Full hierarchical model}\label{sec:hierarchy1}
In what follows, we present the full hierarchical structures of the models for reformulations I and  II, respectively.
The priors on $\theta_{\bm{\gamma}}^\tau$, $\theta_{\bm{z}}^\tau$, $\theta_{\bm{\gamma}}^\kappa$ and  $\theta_{\bm{z}}^\kappa$  for $\mathcal{S} = [0,1]\times[0,1]$ are chosen such that the spatial range and marginal variance vary uniformly in the intervals $[0.01, 1]$ and $[0.01, 5]$, as this should cover all the variability of the considered (scaled) responses. For $\sigma^2_{\varepsilon}$, we specify an inverse gamma distribution $\mathcal{IG}(c = 0.001, d = 0.001)$. Cases $c = d$, with both values approaching zero, are widely used as a weakly informative choice for $\sigma^2_{\varepsilon}$ (see Section 4.4 of \citet{fahrmeirRegression2013}).
Recall that $\gamma(\bm{s}) = \sum^M_{m=1}  \psi_m(\bm{s})\gamma_m$ and $\bm{z}(\bm{s}) = \sum^M_{m=1}  \psi_m(\bm{s})\bm{z}_m$. Moreover, $\bm{\theta_z} = (\theta_{\bm{z}}^\tau, \theta_{\bm{z}}^\kappa)^T$, $\bm{\theta_{z^*}} = (\theta_{z^*}^\tau, \theta_{z^*}^\kappa)^T$ and $\bm{\theta_{\gamma}} = (\theta_{\bm{\gamma}}^\tau, \theta_{\bm{\gamma}}^\kappa)^T$.

The hierarchical structure for $B = 1$ and reformulation I in Section~\ref{sec:re_1} is:
\begin{align*}
y(\bm{s}_i) =& \ \eta(\bm{s}_i) +
\varepsilon_i = \beta_0 + z(\bm{s}_i)^T \bm{\beta} + \gamma(\bm{s}_i) +
\varepsilon_i\ \text{with}  \\
\bm{\gamma} \vert \bm{\theta_z},\bm{\theta_{\gamma}}, \mu_{\bm{z}} &\sim \mathcal{N}(\bm{\mu_{\bm{\gamma} \vert \bm{z}}}, \bm{Q_{\bm{\gamma} \vert \bm{z}}}(\bm{\theta_z},\bm{\theta_{\gamma}})^{-1})\\
\beta_0 &\sim \mathcal{N}(\mu_{\beta_0}, \sigma^2_{\beta_0} ) \\
\bm{\beta} &\sim \mathcal{N}(\bm{\mu_\beta}, \text{diag}(\bm{\sigma^2}_{\bm{\beta}}) ) \\
\bm{\varepsilon} \vert \sigma^2_\varepsilon  &\sim \mathcal{N}(\bm{0}, \sigma^2_\varepsilon \bm{I})\\
\sigma^2_\varepsilon &\sim \mathcal{IG}(0.001, 0.001) \\
\theta_{\bm{\gamma}}^\tau, \theta_{\bm{z}}^\tau &\sim \mathcal{U}(-10, 0) \\
\theta_{\bm{\gamma}}^\kappa, \theta_{\bm{z}}^\kappa &\sim \mathcal{U}(1, 5) \\
\rho &\sim \mathcal{N}(U, 0.05) \\
\mu_{\bm{z}} &\sim \mathcal{N}(\mu_{\mu_{\bm{z}}}, \sigma^2_{\mu_{\bm{z}}}).
\end{align*}

The hierarchical structure for $B = 1$ and reformulation II in Section~\ref{sec:re_2}:
\begin{align*}
y(\bm{s}_i) = \ \eta(\bm{s}_i) +
\varepsilon_i &= \beta_0 + z(\bm{s}_i)^T \bm{\beta} + \gamma(\bm{s}_i) +
\varepsilon_i\ \text{with}  \\
\bm{\gamma} \vert \bm{\theta_z},\bm{\theta_{\gamma}}, \mu_{\bm{z}} &\sim N(\bm{\mu_{\bm{\gamma} \vert \bm{z}}}, \bm{Q_{\bm{\gamma} \vert \bm{z}}}(\bm{\theta_z},\bm{\theta_{\gamma}})^{-1})\\
\beta_0 &\sim \mathcal{N}(\mu_{\beta_0}, \sigma^2_{\beta_0} ) \\
\bm{\beta} &\sim \mathcal{N}(\bm{\mu_\beta}, \text{diag}(\bm{\sigma^2}_{\bm{\beta}}) ) \\
\bm{\varepsilon} \vert \sigma^2_\varepsilon  &\sim \mathcal{N}(\bm{0}, \sigma^2_\varepsilon \bm{I})\\
\sigma^2_\varepsilon &\sim \mathcal{IG}(0.001, 0.001) \\
\theta_{\bm{\gamma}}^\tau, \theta_{\bm{z}}^\tau &\sim \mathcal{U}(-10, 0) \\
\theta_{\bm{\gamma}}^\kappa, \theta_{\bm{z}}^\kappa &\sim \mathcal{U}(1, 5) \\
\rho &\sim \mathcal{PC}(U, 0.05) \\
\mu_{\bm{z}} &\sim \mathcal{N}(\mu_{\mu_{\bm{z}}}, \sigma^2_{\mu_{\bm{z}}}) \\
 \bm{\gamma_z} \vert  \bm{\theta_z}, \mu_{\bm{z}} &\sim N(\mu_{\bm{z}} \bm{1}_M, \bm{Q_z}(\bm{\theta_z})^{-1}).
\end{align*}

\subsection{A novel MGRF prior: multiple covariates case}\label{sec:multivariate}
In this section, the MGRF prior is extended to the case in which multiple covariates are correlated with the spatial random effect. For this, consider the model now includes $B^{*}$ covariates $\bm{z}_1, \cdots, \bm{z}_b, \cdots, \bm{z}_{B^{*}}$ correlated with $\bm{\gamma}$.
The most direct strategy to reduce confounding - considering the prior developed in Section~\ref{sec:newprior} - is to model the whole ($B^{*} + 1$)-variate spatial distribution of $\left( \bm{\gamma}, \;\bm{z}_1, \cdots, \bm{z}_b, \cdots, \bm{z}_{B^{*}} \right)^{T}$. Depending on how large $B^{*}$ is, this strategy could imply a great computational burden, as well as the rather difficult estimation of not only one, but $\frac{B^{*}(B^{*} + 1)}{2}$ correlation parameters.

The prior structure in Section~\ref{sec:newprior} exploits the sparsity of precision matrices of GMRFs, for computational efficiency purposes.
In this section,  an additional source of sparsity is suggested in the multiple covariates case. Namely, we investigate the efficacy of an approach with a ``much lighter'' parameterization, which only requires the estimation of one correlation parameter $\rho$.
In particular,  we investigate the correlation between the confounded $\bm{z}^* = \bm{z}_1 + \cdots + \bm{z}_b + \cdots + \bm{z}_{B^*}$ and $\bm{\gamma}$. Then, we can simply use $\bm{z}^{*}$ as we use $\bm{z}$ in \eqref{eq:cond_dist}. The reasoning underlying our approach is also related to the way spatial confounding arises in Section~\ref{sec:simulationunivariate}, where it is explained in detail.

Our strategy attempts to remove confounding between a linear combination of the covariates $\bm{z}_b$, $b = 1, \ldots, B^{*}$, and the spatial random effect, but does not correct for correlation between the considered covariates.
Nonetheless, if the covariates in $\left( \bm{z}_1, \cdots, \bm{z}_b, \cdots, \bm{z}_{B^{*}} \right)^{T}$ are correlated amongst themselves, the user can choose common solutions against multicollinearity in linear regression models. One possibility is to omit some of the affected covariates or constructing a single combined and easily interpretable covariate from the covariates in question. Another possibility is to use principal component analysis (PCA), with no damage to our general strategy \citep{fahrmeirRegression2013}.

\subsubsection{The full hierarchical model}\label{sec:hierarchy2}
The full hierarchical structure in the multiple covariates case follows closely the structure in Section~\ref{sec:hierarchy1}, but $\bm{z}$ is now replaced by $\bm{z}^*$ in \eqref{eq:cond_dist}. The full hierarchical model can be found in the Supplement~\ref{app:hierarchy2}.

\section{Posterior Evaluation}\label{sec:posterior}

To perform posterior estimation for the full vector of model parameters
$\bm{\vartheta}$, we develop a fully Bayesian approach using MCMC and by separating $\bm{\vartheta}$ into smaller blocks. Of particular relevance is the separation between variables for which we know the corresponding posterior distributions, and those for which the posterior distribution is not available. We will use a Gibbs sampler \citep{gelfand2000gibbs} for the former and derive the corresponding posterior distributions in Section~\ref{sec:gibbs}. For the latter, we use Metropolis Hastings~(MH) \citep{metropolis1953equation} steps, which are explained in more detail in Section~\ref{sec:mh}. Section~\ref{sec:sampler} summarizes the final sampler's structure.

\subsection{Posterior distributions for Gibbs sampler}\label{sec:gibbs}
The priors for $\beta_0$, $\bm{\beta}$, $\bm{\gamma}$, $\sigma_\varepsilon^2$, $\mu_{\bm{z}}$, and additionally $\bm{\gamma_z}$ in reformulation~2, are conjugate
priors such that Gibbs steps are possible. For all parameters that have normal distributed priors, we can use the properties of the product of two normal densities to derive their posterior distribution (see Supplement~\ref{app:ref_2} of \citet{fahrmeirRegression2013}). In what follows, we start by describing the posterior distributions that are identical for both reformulation I and reformulation II, from Section~\ref{sec:re_1} and Section~\ref{sec:re_2}, respectively. Subsequently, we derive the posteriors for the remaining parameters.

\subsubsection{Posterior distributions identical for reformulations I and II}
Let $\tilde{\bm{z}}$ be matrix $\bm{z}$ with first column $\bm{1}_M$. Consider the $(B + 1)\times (B + 1)$ matrices $\bm{A} = \frac{1}{\sigma^2} \tilde{\bm{z}} \bm{\psi}^T \bm{\psi}\tilde{\bm{z}}^T $ and $\bm{B} = \text{diag}\left( \sigma^2_{\beta_0}, {\bm{\sigma}^2_{\bm{\beta}}}^T \right)
^{-1}$, where $\text{diag}$ denotes a diagonal m atrix. Moreover, consider the ($B + 1$)-dimensional vectors $\bm{a} = \bm{\psi}\tilde{\bm{z}}^T\tilde{\bm{z}} \bm{\psi}^T (\bm{\psi}\tilde{\bm{z}}^T\tilde{\bm{z}} \bm{\psi}^T)^{-1}(\bm{y} - \bm{\psi}\bm{\gamma})$, and $\bm{b} = (\mu_{\beta_0}, \bm{\mu_\beta}^T)^T$, where $\bm{y} = (y(\bm{s}_1), \ldots, y(\bm{s}_n))^T$. Then,
\begin{equation}\label{eq:posterior}
 (\beta_0, \bm{\beta}^T)^T \vert \cdot \sim \mathcal{N} \left( (\bm{A} + \bm{B})^{-1}(\bm{A}\bm{a} + \bm{B}\bm{b}), (\bm{A} + \bm{B})^{-1} \right).
\end{equation}

Consider the single covariate case and Section~\ref{sec:hierarchy1}.
The posterior for $\mu_{\bm{z}}\vert \cdot$ follows the same structure as \eqref{eq:posterior}, but we instead have the scalars $A = \bm{1}_M^T\bm{Q_Z}\bm{1}_M$, $B = \frac{1}{\sigma^2_{\mu_{\bm{z}}}}$, and the $M$-size vectors $a = \bm{1}_M^T \bm{I}_M \bm{z}$ and $b = \mu_{\mu_{\bm{z}}}$,  with $\bm{z}$ replaced by $ \bm{z^*}$ in the multiple covariates case.

In the case of $\sigma_\epsilon^2$, the prior is inverse gamma distributed, which combined with the normal distributed of the response leads to a posterior distribution (see p. 229 of \citet{fahrmeirRegression2013})

$$\sigma_\varepsilon^2 \sim \mathcal{IG}\left( 0.001 + \frac{n}{2}, 0.001 + \frac{1}{2}(\bm{y} - \bm{\eta})^T(\bm{y} - \bm{\eta})\right),$$
where $\bm{\eta} = (\eta(\bm{s}_1), \ldots, \eta(\bm{s}_n))^T$ and, according to Section~\ref{sec:hierarchy1} and Section~\ref{sec:hierarchy2}.

\subsubsection{Posterior distributions not identical for reformulations I and II}
In reformulation I, $\bm{\gamma}\vert \cdot$ once again follows the structure in \eqref{eq:posterior}, but with $M \times M$ matrices $\bm{A} = \frac{1}{\sigma^2} \bm{\psi}^T  \bm{\psi}$ and $\bm{B} = \bm{Q_{\bm{\gamma} \vert \bm{z}}}^{-1}$ and $M$-size matrices $\bm{a} = \bm{y} - \beta_0\bm{1}_n - \bm{\psi}\bm{z}^T\bm{\beta}$ and $\bm{b} = \bm{\mu_{\bm{\gamma} \vert \bm{z}}}$. In the multiple covariates case, $\bm{B} = \bm{Q_{\bm{\gamma} \vert \bm{z^*}}}^{-1}$ and $\bm{b} = \bm{\mu_{\bm{\gamma} \vert \bm{z^*}}}$.

In reformulation II, $\bm{\gamma}\vert \cdot$ is sampled according to $M \times M$ matrices $\bm{A} = \frac{1}{\sigma^2} \bm{\psi}^T  \bm{\psi}$ and $\bm{B} = \bm{Q}_{\bm{\gamma}}^{-1}$ and size-$M$ matrices $\bm{a} = \bm{y} - \beta_0\bm{1}_n - \bm{\psi}\bm{z}^T\bm{\beta} - \rho\, \bm{L}_{\bm{\gamma}}^{-T}  \bm{L}_{\bm{z}}^{T}( \bm{z} -  \bm{\gamma_z})$ and $\bm{b} = \bm{0}$. In the multiple covariates case, $\bm{a} = \bm{y} - \beta_0\bm{1}_n - \bm{\psi}\bm{z}^T\bm{\beta} - \rho\, \bm{L}_{\bm{\gamma}}^{-T}  \bm{L}_{\bm{z}^*}^{T}( \bm{z}^* -  \bm{\gamma_{z^*}})$.

Finally, in reformulation II, $ \bm{\gamma_z}\vert \cdot$ is sampled according to the $M \times M$ matrices $\bm{A} = (\rho\, \bm{L}_{\bm{\gamma}}^{-T} \bm{L}_{\bm{z}}^{T})^{T} \bm{Q}_{\bm{\gamma}} \rho\, \bm{L}_{\bm{\gamma}}^{-T} \bm{L}_{\bm{z}}^{T}$, $\bm{B} = \bm{Q_z}$, $\bm{a} =  \rho\, \bm{L}_{\bm{\gamma}}^{-T} \bm{L}_{\bm{z}}^{T} \bm{z} - \bm{\gamma}$ and $\bm{b} = \mu_{\bm{z}} \bm{1}_K$, with all $\bm{z}$ and $ \bm{\gamma_z}$ replaced by $ \bm{z^*}$ and  $\bm{\gamma_{z^*}}$, respectively, in the multiple covariates case.

\subsection{Posterior estimation for MH steps}\label{sec:mh}
We now turn to parameters $\bm{\theta_\gamma}$, $\bm{\theta_z}$, and $\rho$, for which MH steps are used.
First, parameter $\rho$ is re-parameterized using the general Fisher's $z$-transformation \citep{fisher1958statistical} such that
\begin{equation*}
  \rho = \frac{\exp(\rho^*) - 1}{\exp(\rho^*) + 1} \quad \text{and} \quad \rho^* = \log \left( \frac{1 + \rho}{1 - \rho} \right).
\end{equation*}
We use the change of variable theorem to correct the full conditional of $\rho$ for this re-parameterization (see Theorem B.1 on Supplement~B.1 of \citet{fahrmeirRegression2013}).
Then, as parameters $\bm{\theta_\gamma}$, $\bm{\theta_z}$, and
$\rho^*$ are highly correlated, we sample them in one block in an Metropolis-Hastings~(MH) step using the robust
adaptive MH method from \citet{vihola2012robust} with Student's t-distributed proposal densities. This algorithm estimates the shape of the target distribution and simultaneously coerces the acceptance rate. Here, we use the typically desired acceptance rate of $23.4\%$ for multidimensional settings \citep{roberts1997weak}.

\subsection{Sampler}\label{sec:sampler}
With a focus in the one covariate case, the final MCMC scheme is outlined in Algorithm~\ref{algo:mcmc-outline}. It follows similarly for the multiple covariates case.

\begin{algorithm}[ht!]
\caption{Outline of the MCMC sampler for the one covariate case.}\label{algo:mcmc-outline}
\SetAlgoLined
\KwData{Starting values $\bm{\vartheta}_0$. Number of iterations $T$.}
t = 0\;
\While{$t <  T$}{
		Generate move from $p(\mu_{\bm{z}} \vert \cdot)$\;
		Perform MH step for $\bm{\theta_z}$\;
		(In reformulation II, generate  move from $p(\bm{\gamma_z} \vert \cdot)$\;
		Generate move from $p(\sigma_\varepsilon^2 \vert \cdot)$\;
		Generate move from $p((\beta_0,\bm{\beta}^T)^T \vert \cdot)$\;
		Perform MH step for $\lbrace \bm{\theta_\gamma}, \rho \rbrace$\;
		Generate  move from $p(\bm{\gamma} \vert \cdot)$\;
		t += 1\;
}
\end{algorithm}

\newpage

\section{Simulation Study}\label{sec:simulation}
The aim of this simulation section is three-fold. First, we want to understand the sources and consequences of spatial confounding in spatial models.  Second, it is in our interest to investigate to which extent the priors developed in Section \ref{sec:newprior} and \ref{sec:multivariate} reduce the bias of regression coefficients when spatial confounding is present, and third, we evaluate whether our model is able to recover the true $\rho$. For this, we conduct two simulation studies, corresponding to the univariate and a multivariate cases. In addition, the coverage rates of the considered regression coefficients are provided in Supplement~\ref{app:coverage_rate}.

We compare four different specifications of the model in \eqref{eq:model_equation}: (i) a non-spatial model; i.e., one that excludes $\gamma(\bm{s})$, (ii) a spatial model with the base prior, (iii) a RSR model, and (iv) a spatial model with the MGRF prior. We sample $\gamma(\bm{s})$ in RSR using conditioning kriging (see Supplement~\ref{app:rsr_sampler}). Moreover, we use $\mathcal{S} = [0,1]\times[0,1]$ and a finite elements mesh with $M = 523$ nodes (see Supplement~\ref{app:crps}).

The results of the simulation studies for simulation study 1 and simulation study 2 are presented in Section~\ref{sec:simulationunivariate} and Section~\ref{sec:simulation_multivariate}, respectively. The results for each scenario are based on $N = 50$ replicates.
Inference is based on
12000 MCMC samples with burn-in of 6000 and thinning of 1. We use reformulation I from Section~\ref{sec:re_1}, as the true values of $\rho$ are far from the boundaries.

\subsection{Simulation study 1: sources of confounding and performance of the prior in the one covariate case}\label{sec:simulationunivariate}

\subsubsection{Data generation}

The data is generated as follows: the covariate $z(\bm{s})$ and the spatial effect $\gamma(\bm{s})$ are GRFs with mean zero generated with the SPDE-approach and with marginal variance and smoothness equal to one, i.e.,  $\sigma^2_{\bm{z}} = \sigma^2_{\bm{\gamma}} = 1$ and $\nu^2_{\bm{z}} = \nu^2_{\bm{\gamma}} = 1$, respectively. We consider nine pairs of spatial ranges $(r_{\bm{\gamma}},r_{\bm{z}})^T$ where $r_{\bm{\gamma}},r_{\bm{z}} \in \lbrace 0.1, 0.5, 0.9 \rbrace$. Furthermore, $\sigma^2_\varepsilon = 0.1$ and $ (\beta_0, \beta) = (\beta_0, \beta_1) = (-1.5, 1)$.
We consider five scenarios, each with a different $\rho$: (1) $\rho = 0$; (2) $\rho = 0.3$; (3) $\rho = - 0.3$; (4) $\rho = 0.7$; (5) $\rho = - 0.7$.

Scenario~1 sheds some light on potential issues with overfitting and on whether fixed and spatial random effects can be confounded even when they are independent. Moreover, $\rho = 0$ and $\mu_{\bm{z}} = 0$ implies $\bm{z}^T\bm{\gamma} = 0$, i.e., the covariate is orthogonal to the spatial random effect on the full spatial support of the system and, thus, this dataset follows a RSR model.

\subsubsection{Prior elicitation}
We use the priors $\mu_{\bm{z}} \sim \mathcal{N}(0, 0.1^2)$ and $(\beta_0, \beta_1)^T \sim \mathcal{N}(\bm{0}, \mbox{diag}(100^4, 100^2))$.
The values of $U$ in the shape of the PC-prior (see \eqref{eq:lambda}) vary between scenarios, and represent a trade-off between the accuracy of the posterior mean of $\beta_1$ over 50 replications and the dispersion of these posterior means, as explained in more detailed below. Additional details on how to set-up the PC-prior are provided in Supplement~\ref{app:pc_more}.

 \subsubsection{Results}
\begin{figure}[b!]
    \begin{minipage}{0.44\textwidth}
      \centering
      \includegraphics[width=\textwidth]{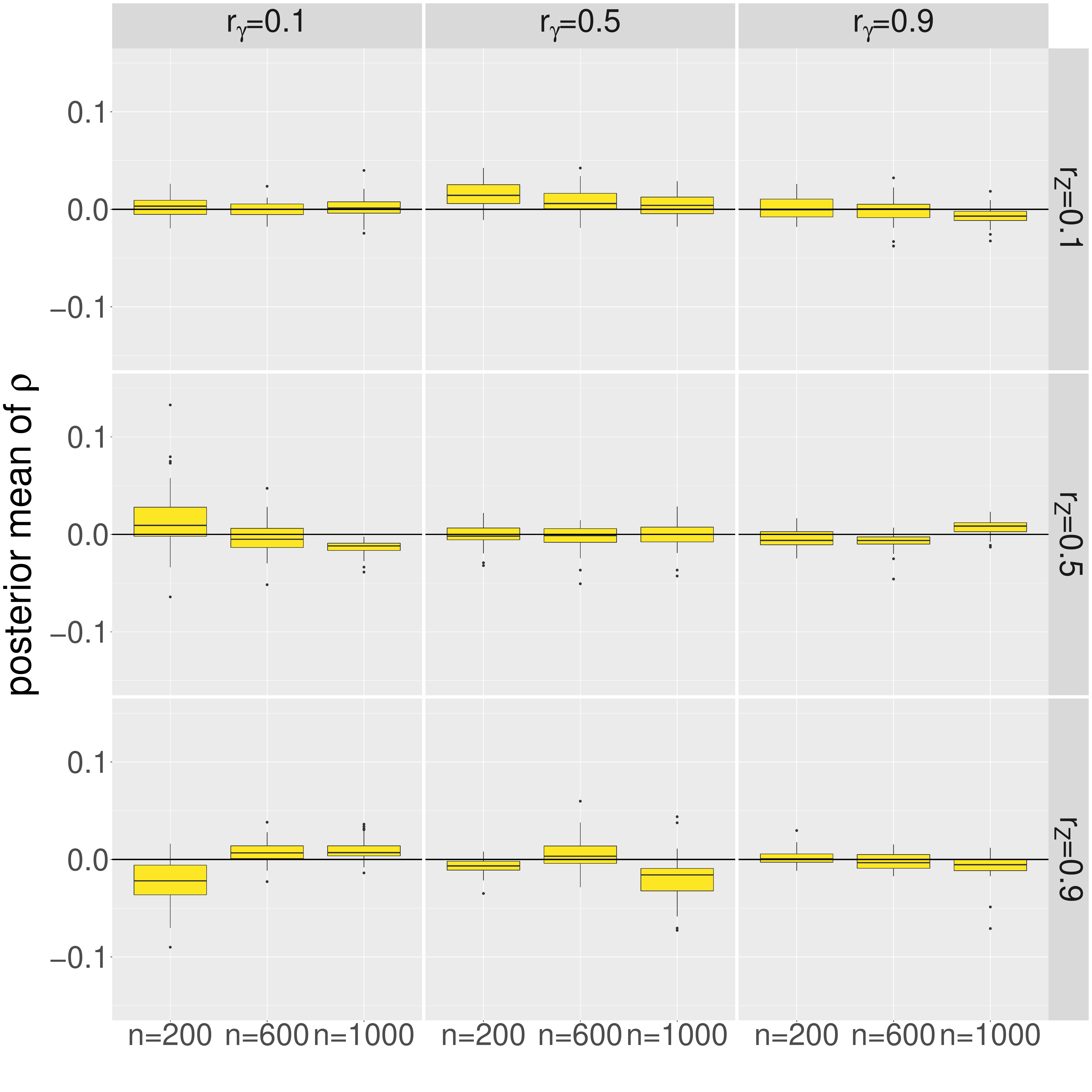}
    \end{minipage}
    \begin{minipage}{0.55\textwidth}
      \centering
    \includegraphics[width=\textwidth]{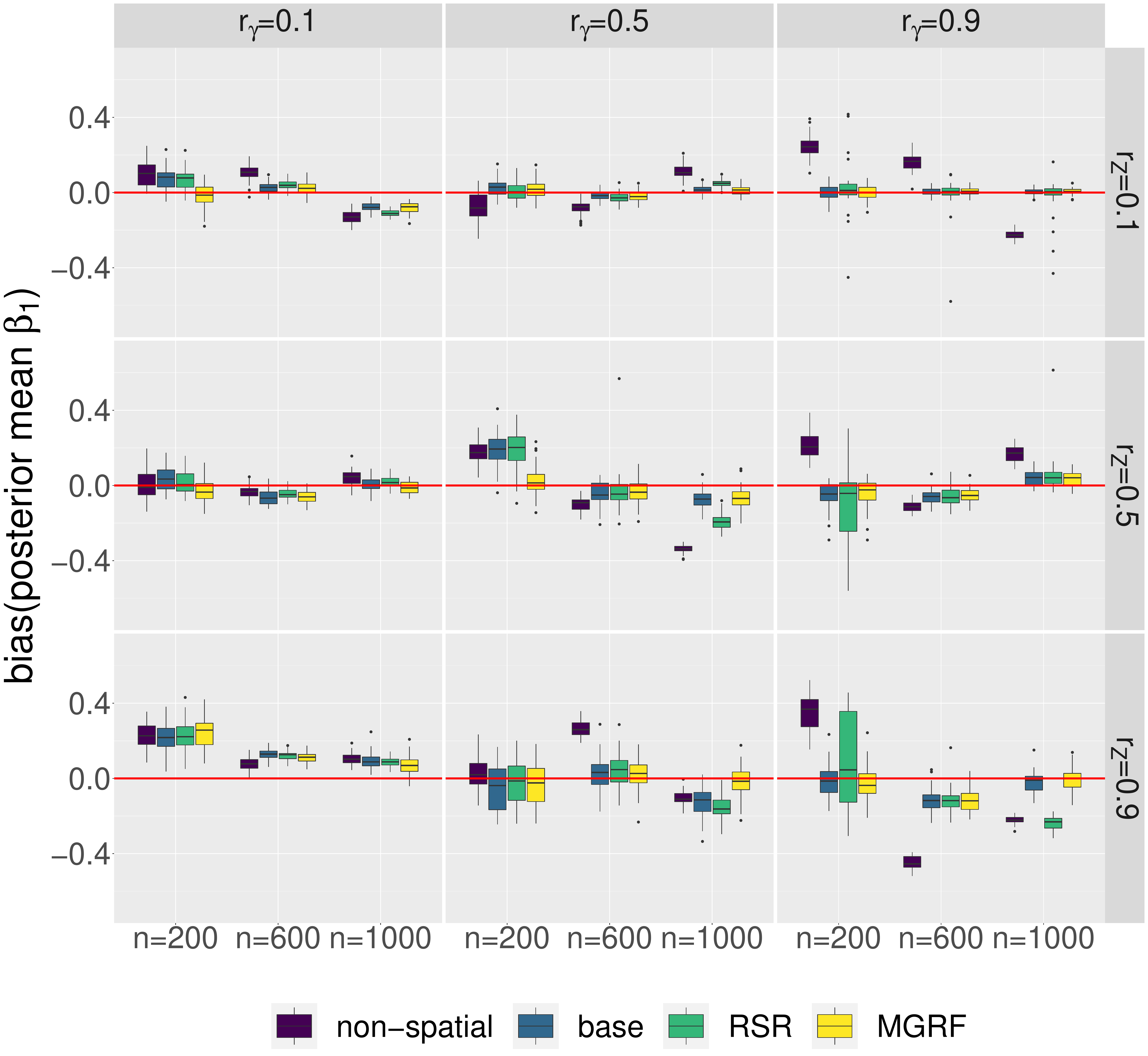}
    \end{minipage}
    \caption{Results for Scenario~1. From left to right: posterior mean of $\rho$ for spatial model with MGRF prior and bias of the posterior mean of $\beta_1$ for 50 replicates. The red and black lines indicate a value of 0.}\label{fig:rhozero}
\end{figure}

In Figure \ref{fig:rhozero}, corresponding to Scenario~1, a PC-prior prevents overfitting by keeping the values $\rho$ close to zero. The values of $\rho$ only diverge from zero in scenarios where this reduces the bias of $\beta_1$, e.g., when $r_{\bm{\gamma}} = 0.5$ and $r_{\bm{z}} = 0.9$. In general, when $r_{\bm{z}}\geq r_{\bm{\gamma}}$, the estimates of $\beta_1$ are slightly biased, despite $\bm{z}$ and $\bm{\gamma}$ being independent. The non-spatial model and RSR tend to perform worse than the remaining two models.

\begin{figure}[t!]
  \centering
    \includegraphics[width=\textwidth]{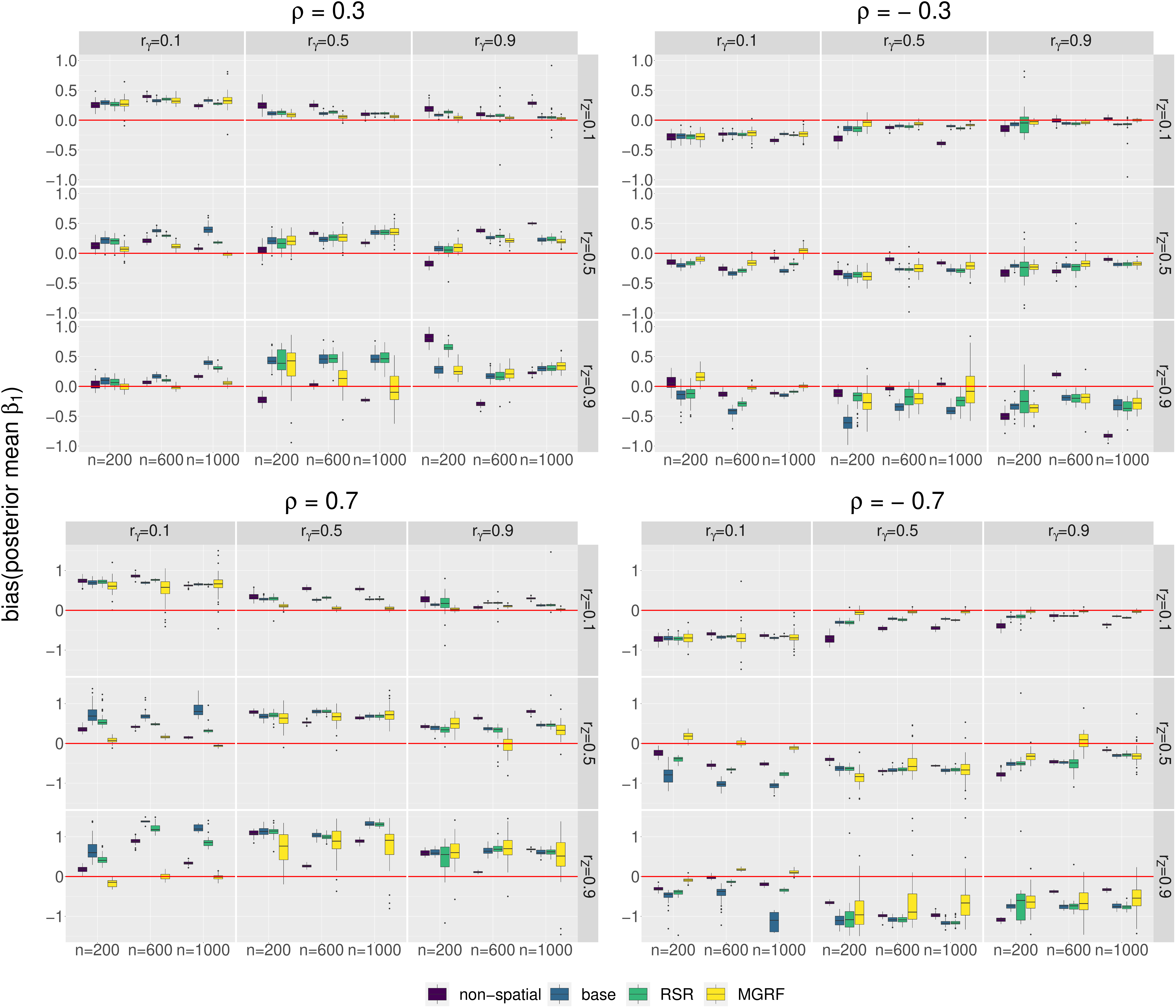}
    \caption{From left to right, top to bottom: bias of the posterior mean of $\beta_1$ for 50 replications in Scenarios 2, 3, 4 and 5, respectively. Note that the scale of the bias is twice as large when $|\rho| = 0.7$.}\label{fig:rhonotzero}
\end{figure}

We now turn to the scenarios where $\bm{z}$ and $\bm{\gamma}$ are not independent, i.e., the data is generated with $\rho \neq 0$. In Figure \ref{fig:rhonotzero}, we observe that spatial confounding leads to biased regression coefficients when $r_{\bm{z}} \geq r_{\bm{\gamma}}$. As a consequence, our prior performs considerably better for cases with $r_{\bm{z}} > r_{\bm{\gamma}}$. When $r_{\bm{\gamma}} = r_{\bm{z}}$, as expected, the effects cannot be separated well, as both $\bm{z}$ and $\bm{\gamma}$ have the same spatial structure. These results are in line with \citet{paciorek2010importance} and \citet{page2017estimation}.
 Moreover, there is a tendency to overestimate $\beta_1$ for $\rho > 0$ and to underestimate $\beta_1$ for $\rho < 0$. For symmetric values of the true $\rho$, the bias of $\beta_1$ also behaves approximately symmetrically, e.g., Scenario~2 compared to Scenario~3.

\begin{figure}[t!]
  \centering
  \includegraphics[width=\textwidth]{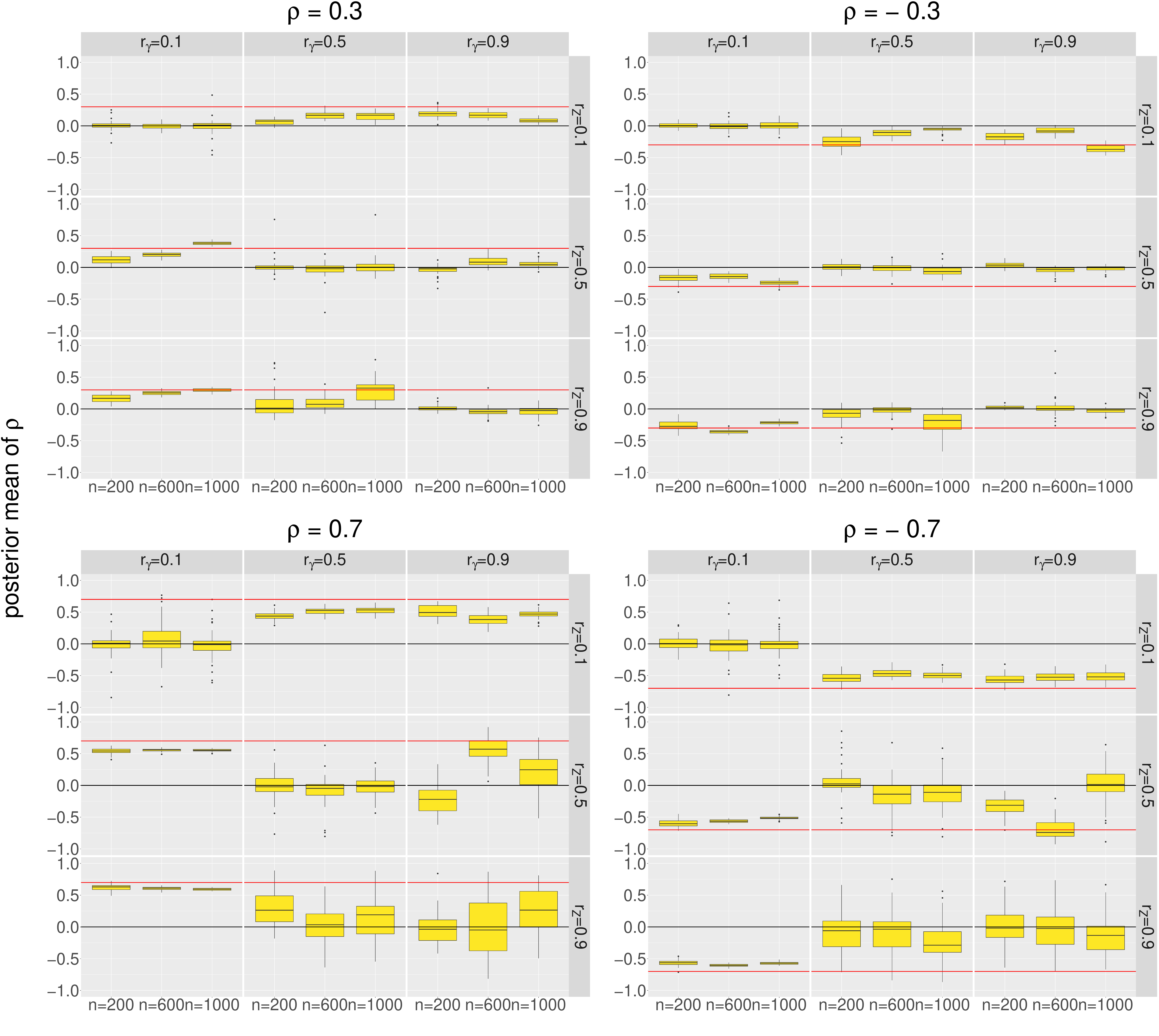}
  \caption{From left to right, top to bottom: posterior mean of $\rho$ for 50 replications in Scenarios 2, 3, 4 and 5, respectively. The red line represents the true value and the black line indicates a value of 0.}\label{fig:rhoandnotzero}
\end{figure}

In Figure \ref{fig:rhoandnotzero}, the PC-prior for $\rho$ prevents overfitting by keeping $\rho$ close to zero when it cannot separate the effects of $\bm{z}$ and $\bm{\gamma}$, or when the bias is small to begin with. When $r_{\bm{z}} > r_{\bm{\gamma}}$, the posterior mean of $\rho$ over the 50 replications gets very close to the true value of $\rho$. By the very nature of a PC-prior, in scenarios that are harder to estimate the posterior mean of $\rho$ tends to cycle between iterations: it is either close to 0 or close to the true value. This can lead to very broad boxplots for the posterior mean of $\rho$ and, consequently, also of $\beta_1$. This effect can be reduced by using a more conservative PC-prior, i.e., a lower value for $U$.

 All in all, the MGRF prior always performs better or similarly well, in terms of the bias of the posterior mean of $\beta_1$, than the base prior or the RSR. In certain cases, namely for small sample sizes, a non-spatial regression model performs better than any of the spatial models. All spatial models reach similar values for the continuous ranked probability score (CRPS), but predictions for the non-spatial model are noticeably worse (see Supplement~\ref{app:crps}).

A brief report on the efficiency of the sampler for the different models is provided on Supplement~\ref{app:rsr_sampler}. Coverage rates are presented in Supplement~\ref{app:coverage_rate}.

\subsection{Simulation study 2: performance of the prior in the multiple covariates case}\label{sec:simulation_multivariate}

 In simulation study 2, we consider multiple covariates $\bm{z}(\bm{s}) = (z_1(\bm{s}), z_2(\bm{s}))^T$ and test the prior structure suggested in Section \ref{sec:multivariate}. We consider three scenarios, with different values for $\bm{\rho} = (\rho_1, \rho_2)^T$: (1) $(\rho_1, \rho_2) = (0, 0.3)$; (2) $(\rho_1, \rho_2) = (0.7, 0.3)$; (3) $(\rho_1, \rho_2) = (- 0.3, 0.7)$.

\subsubsection{Data generation}

 The data is generated as follows: covariates $z_1(\bm{s})$, $z_2(\bm{s})$ and the spatial effect $\gamma(\bm{s})$ are GRFs with mean zero generated with the SPDE-approach and with marginal variance and smoothness equal to one; i.e., $\sigma^2_{\bm{z}_1} = \sigma^2_{\bm{z}_2} = \sigma^2_{\bm{\gamma}} = 1$ and $\nu^2_{\bm{z}_1} = \nu^2_{\bm{z}_2} = \nu^2_{\bm{\gamma}}$, respectively. For the sake of simplicity, we fix $r_{\bm{z}_1} = 0.5$ and consider nine pairs of $(r_{\bm{\gamma}},r_{\bm{z}_2})$ where $r_{\bm{\gamma}},r_{\bm{z}_2} \in \lbrace 0.1, 0.5, 0.9 \rbrace$. Furthermore, $\sigma^2_\varepsilon = 0.1$, $\bm{\beta} = (\beta_1, \beta_2)^T$  and $(\beta_0, \beta_1, \beta_2) = (-1.5, 1, -0.5)$.

\subsubsection{Prior elicitation}
The choice of $U$ follows the same strategy as in Section~\ref{sec:simulationunivariate}.
We consider the priors \newline ${\mu_{\bm{z}}}^{*} \sim \mathcal{N}(0, 0.1^2)$ and $(\beta_0, \beta_1, \beta_2)^{T} \sim \mathcal{N}(\bm{0}, \mbox{diag}(100^4, 100^2, 100^2))$.

\subsubsection{Results}

Here, we compare the sum of the absolute bias of the posterior mean of $\beta_1$ and absolute bias of the posterior mean of $\beta_2$ for all replications:
\begin{align*}
  \mbox{bias}_{B^*} = |\mbox{(bias(posterior mean of }\beta_1)| + |\mbox{(bias(posterior mean of }\beta_2)|.
\end{align*}
The results are presented in Figure~\ref{fig:multivariate_bias} and the bias for individual coefficients can be found in Supplement~\ref{app:beta1beta2_multivariate}.

 \begin{figure}[t]
  \centering
    \includegraphics[width=\textwidth]{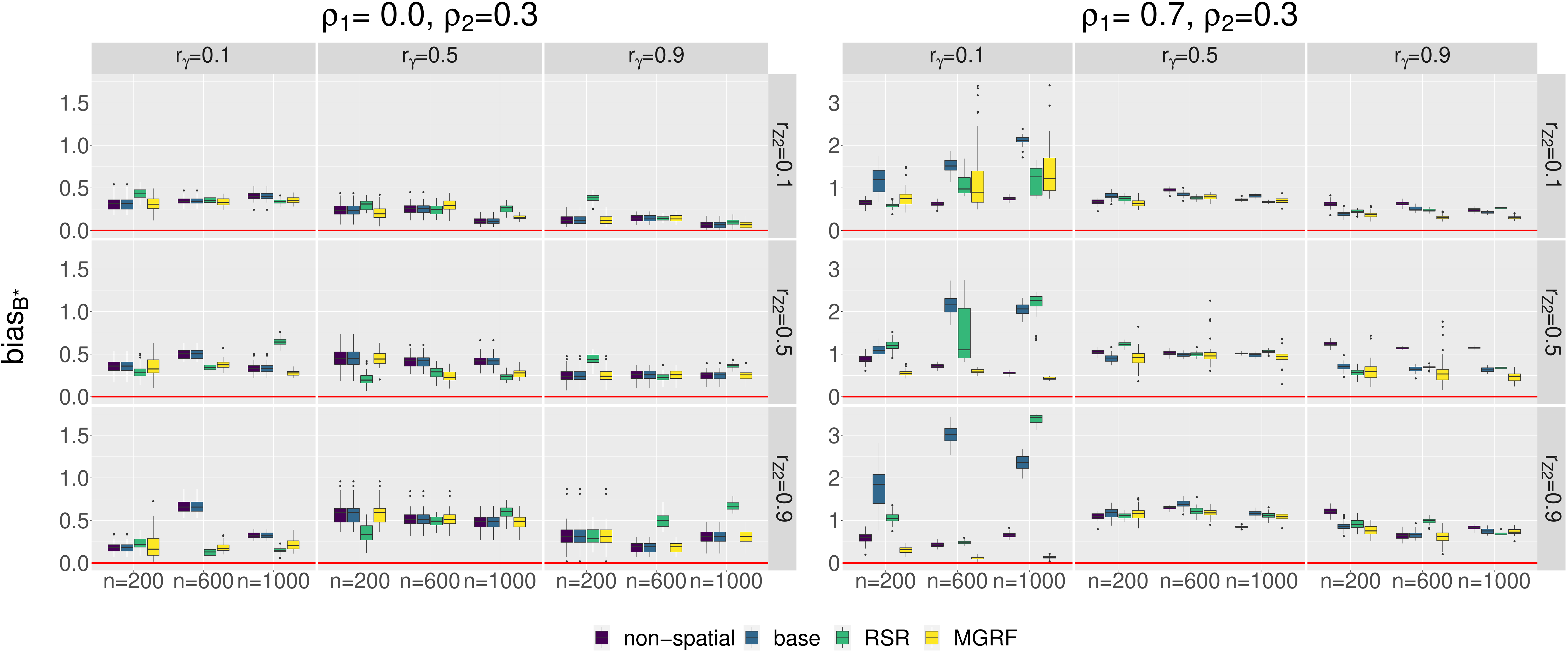}
    \caption{From left to right: the sum of the absolute bias of the posterior mean of $\beta_1$ and $\beta_2$ for 50 replications in Scenario 1 and Scenario~2, respectively. Note that the scale of the bias is  twice as large in Scenario~2.}\label{fig:multivariate_bias}
\end{figure}

\begin{figure}[b!]
  \centering
    \includegraphics[width=\textwidth]{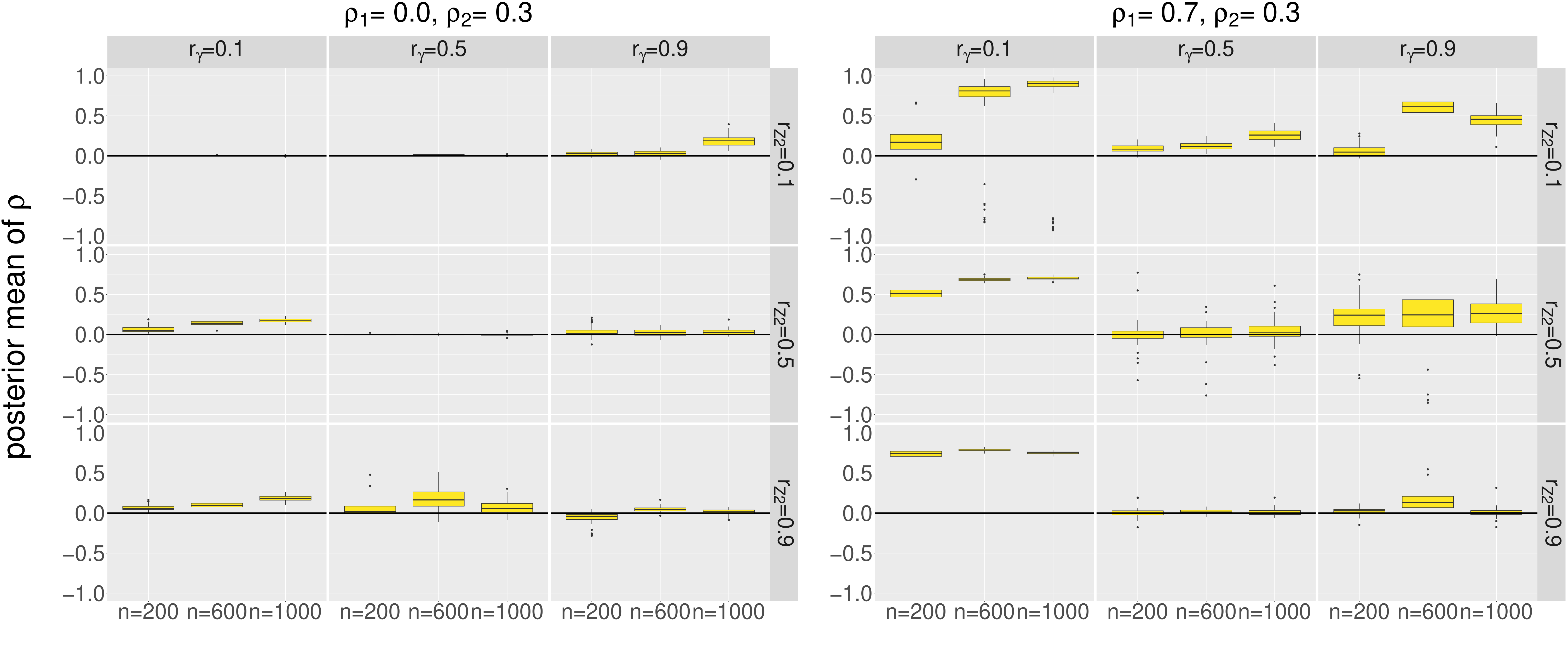}
    \caption{From left to right: posterior mean of $\rho$ in Scenario~1 and Scenario~2, respectively.}\label{fig:rhomultivariate}
\end{figure}

According to Figure \ref{fig:rhomultivariate}, in Scenario~1, $\rho$ tends to be conservative and only deviates from zero for cases with $r_{\bm{z}_2} \geq r_{\bm{\gamma}}$, in which we also expect a larger bias of the regression coefficient $\beta_2$. Thus, as expected, this scenario behaves similarly to the one covariate case.

In Scenario~2, we observe that when $\bm{z}_2$ has a low spatial range, and thus ``low signal'', we get values of the posterior mean of $\rho$ that are close to the true value of $\rho_1$, i.e., 0.7. When $\bm{z}_2$ also has high signal ($r_{\bm{z}_2} > 0.1$), the posterior mean of $\rho$ stays systematically at high values (close to 0.7), leading to a great bias reduction compared to RSR or the marginal prior.

\begin{figure}[tb]
      \centering
      \includegraphics[width=\textwidth]{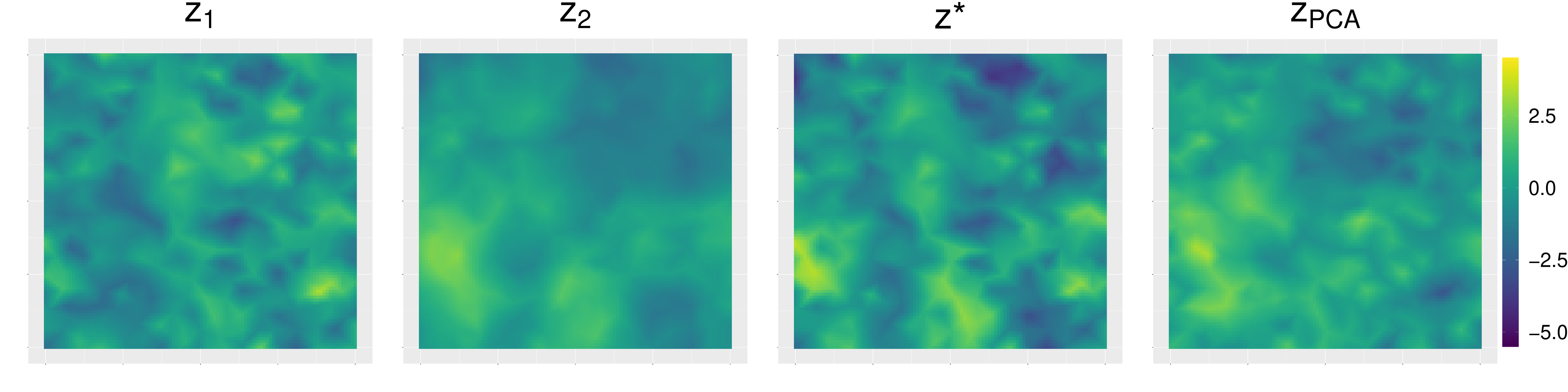}
  \caption{Visual aid for understanding the prior for multiple covariates, represented on $\mathcal{S} = [0, 1] \times [0, 1]$. On the left, values of $\bm{z}_1$ with a spatial range of 0.1 and marginal variance 1. On the center-left, values of $\bm{z}_2$ with a spatial range 0.5 and marginal variance 1. On the center-right, values $\bm{z}^*$. On the right, $\bm{z}_{PCA}$.}\label{fig:intuitive_multivariate}
\end{figure}

Essentially, following the results from Section \ref{sec:simulationunivariate}, we say spatial covariates have a larger signal the larger their spatial range is. If the covariate's spatial range is larger than the unobservable spatial effect's, the covariate \textit{dominates}
 and spatial confounding arises. Similarly, with the sum $\bm{z}^* = \bm{z}_1 + \cdots + \bm{z}_b + \cdots + \bm{z}_{B^*}$ we expect to recover an overall spatial ``dominance'' in the linear predictor. Figure \ref{fig:intuitive_multivariate} provides a visual aid. The covariates $\bm{z}_1$ and $\bm{z}_2$ the figure have a spatial range of 0.1 and 0.5, respectively. The marginal variance is 1 for both.
 The covariate
  $\bm{z}_2$ has a larger spatial range than $\bm{z}_1$ and the resulting covariate $\bm{z}^*$ captures the spatial behavior of both covariates, while $\bm{z}_2$ dominates, i.e., the spatial range of $\bm{z}_2$ dominates. We compare this with the result of the PCA of $(\bm{z}_1, \bm{z}_2)^T$, considering only the first principal component. This helps introducing Scenario~3. The covariate resulting from PCA, $\bm{z}_{PCA}$, is smoother, since part of the effect of $\bm{z}_1$ - the covariate with the smallest spatial range - is excluded from the first component of the PCA.

\begin{figure}[ht!]
    \begin{minipage}{0.45\textwidth}
      \centering
      \includegraphics[width=\textwidth]{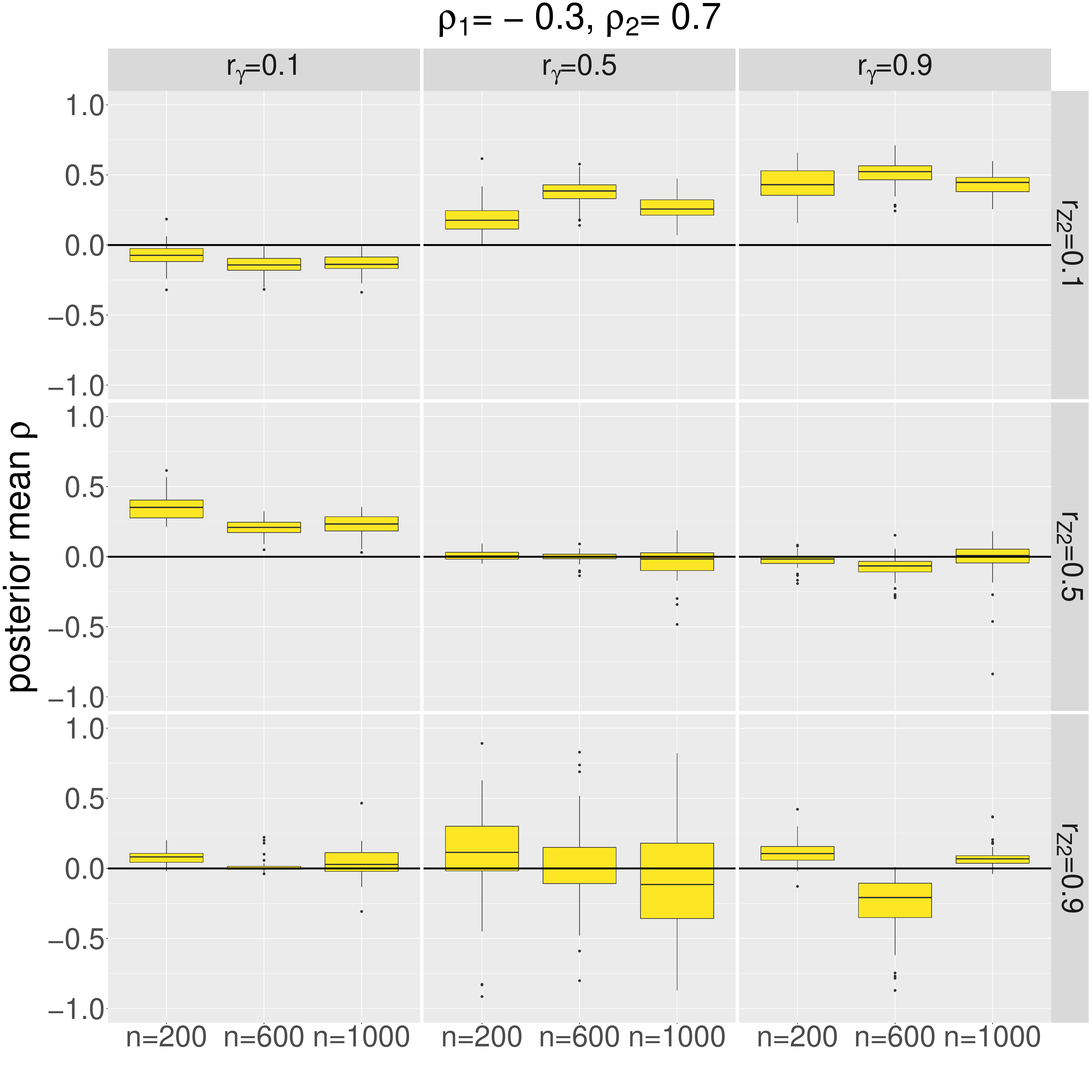}
    \end{minipage}
    \begin{minipage}{0.54\textwidth}
      \centering
    \includegraphics[width=\textwidth]{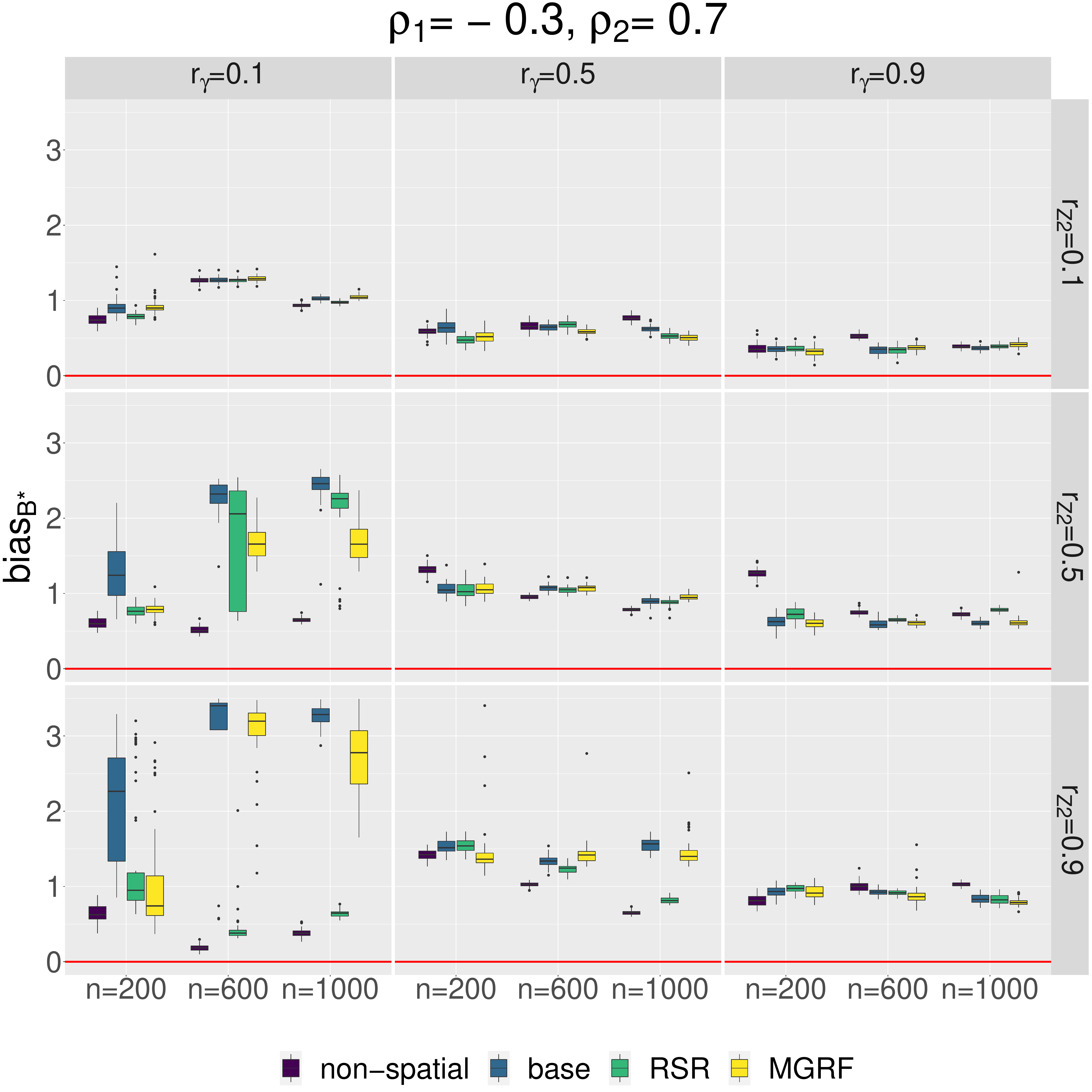}
    \end{minipage}

    \begin{minipage}{0.45\textwidth}
      \centering
    \includegraphics[width=\textwidth]{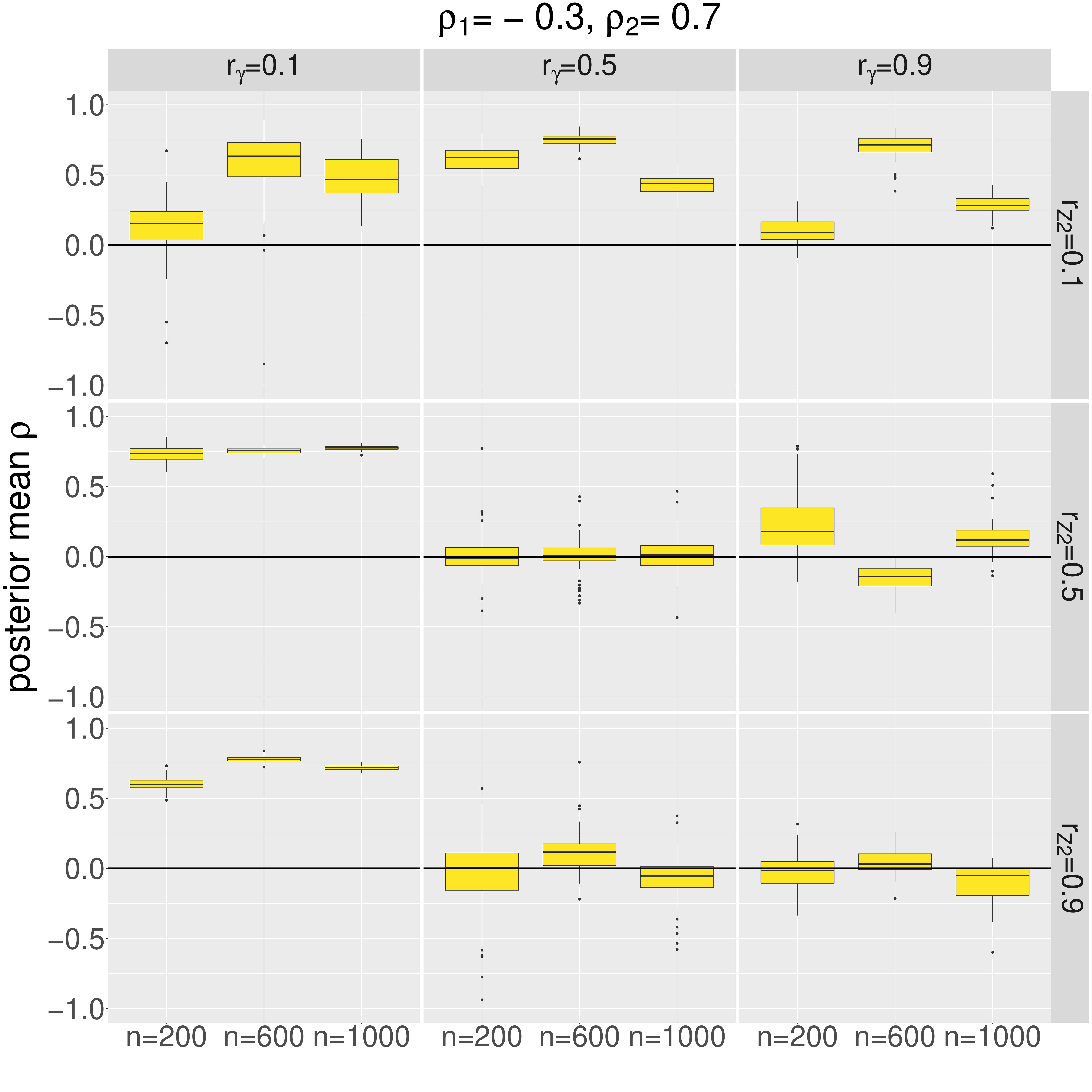}
    \end{minipage}
    \begin{minipage}{0.54\textwidth}
      \centering
    \includegraphics[width=\textwidth]{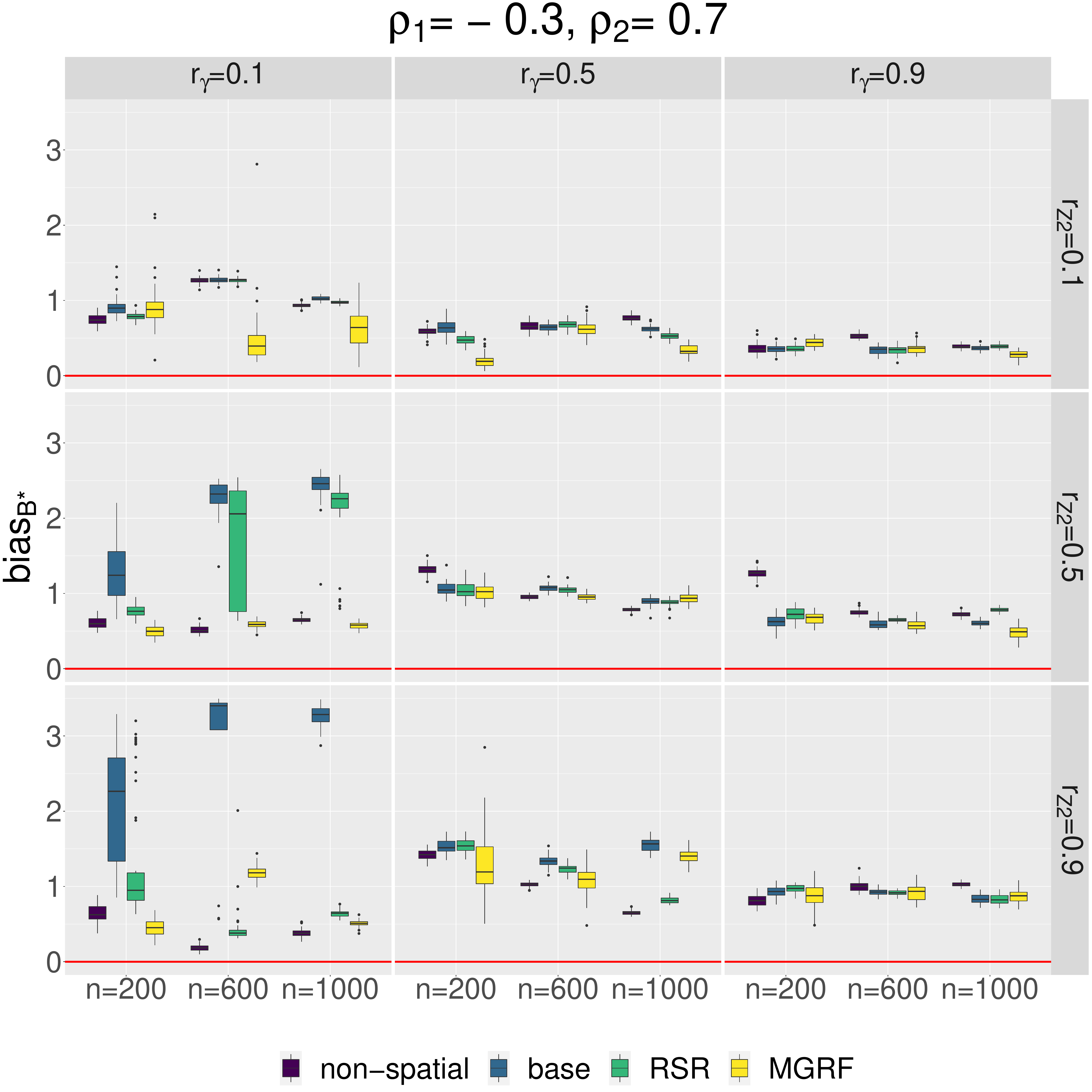}
    \end{minipage}
    \caption{Results for Scenario~3. From left to right, top to bottom: posterior mean of $\rho$ and bias of the posterior mean of $\beta_1$ for 50 replicates, for the strategy from Section~\ref{sec:multivariate} and PCA, respectively.}\label{fig:rhomultivariate3}
\end{figure}

Scenario~3 behaves very differently from Scenario~1 and Scenario~2, and it is presented separately. Since the correlations of $\bm{z}_1$ and $\bm{z}_2$ with the spatial effect have opposite signs, the PC-prior for $\rho$ has difficulty capturing the correlation between $\bm{z}^{*} = \bm{z}_1 + \bm{z}_2$ and the spatial effect. Thus, the PC-prior for $\rho$ tends to be conservative, i.e., stays close to 0. Additionally, we test a PCA of $(\bm{z}_1, \bm{z}_2)^{T}$ with only one principal component. The results are presented in Figure \ref{fig:rhomultivariate3}, together with the results for our original method. The PCA method leads to a much smaller bias, especially when $r_{\bm{z}_2} > r_{\bm{\gamma}}$. The reason for this is that the PCA seems to partly exclude the effect of $\bm{z}_1$ and essentially only accounts for the correlation between $\bm{z}_2$ and $\bm{\gamma}$: the bias of $\beta_2$ is improved to no detriment to $\beta_1$ (see Supplement~\ref{app:beta1beta2_multivariate}).

However, the previous results do not imply PCA performs better in every scenario. In Scenario~2, when using PCA we still get considerable improvements over RSR, but the results are worse than the results presented in this section (see Supplement~\ref{app:pca_scenario2}). The reason is that, once again, PCA tends to focus on one covariate, but in this case the the correlation between the two covariates and the spatial random effect has the same signal, and using $\bm{z}^{*}$ leads to better results.

\subsection{Concluding remarks}
To conclude, our prior structure reduces bias significantly, while in certain scenarios it is still far from reaching the desirable bias of (about) zero. Further work needs to be done on the MGRF prior for the case of multiple covariates, so that it can flexibly adapt to different data structures, i.e., a model that inherently addresses the contrast between the data types in Scenarios 1 and 2, and the one in Scenario~3 (the strategy from Section \ref{sec:multivariate} as opposed to PCA). In general, the advantage of a straightforward strategy such as RSR becomes clearer in the multiple covariates case, where currently our model still requires considerable tuning by the user. Nonetheless, the user should be aware that RSR can aggravate bias compared to the base model, as can be observed in Section~\ref{sec:simulation_multivariate}. Finally, in future work we should also consider the effect of different marginal variances of the GRF on spatial confounding.

\section{Application to daily precipitation in Germany}\label{sec:application}

In this section, we analyze stylized facts of daily precipitation in Germany using data from the  German Meteorological Institute (Deutscher Wetterdienst, DWD)\footnote{\url{https://www.dwd.de}}. The website also contains detailed information on elevation and climatic variation at each weather station, which we can use as covariates in our geospatial model.
We study two different days, January 1 and March 1, 2000. These days demonstrate the range of possible effects that using a MGRF prior has on the posterior distribution of $\bm{\beta}$.

We consider the model in \eqref{eq:model_equation}. Variable $y(\bm{s}_i)$ is the \textit{standardized} amount of precipitation in milliliters, at weather station $\bm{s}_i \in \mathcal{S}$ and $\mathcal{S}$ represents the whole of Germany. Covariate
$z_1(\bm{s}_i)$ is the standardized elevation in meters and $z_2(\bm{s}_i)$ is the standardized minimum temperature in degrees Celsius.
We standardize all variables because the effects of different marginal variances of the GRFs was not analyzed in Section~\ref{sec:simulation}.
Intuitively, one would expect elevation to be associated with a positive effect ($\beta_1 > 0$) on the mean of daily precipitation and minimum temperature to be associated with a negative effect ($\beta_2 < 0$).

 Throughout the application, we consider the priors $\mu_{\bm{z}} \sim \mathcal{N}(0, 1^2)$  and \linebreak $(\beta_0, \beta_1,  \beta_2)^T \sim \mathcal{N}(\bm{0}, \mbox{diag}(100^4, 100^2, 100^2))$. Results do not seem overly sensitive to different values of $U$ in the interval $[0.7,0.9]$ and we choose $U = 0.8$. The remaining priors are the ones presented in Section~\ref{sec:hierarchy1} and Section~\ref{sec:hierarchy2}, for the single and multiple covariate cases, respectively.

 The results are based on data from 280 DWD weather stations.
 The mesh used has $278$ nodes and the domain is re-scaled such that $\mathcal{S} \subset [0,1] \times [0,1]$. The resulting mesh is presented in Supplement~\ref{app:mesh}.
We test five models with different specifications for $\gamma(\bm{s})$, as done in the simulation studies from Section~\ref{sec:simulation}. Concretely: (i) $\gamma(\bm{s})$ is excluded from the model, i.e., non-spatial model, (ii) $\gamma(\bm{s})$ has a base prior, (iii) $\gamma(\bm{s})$ has a MGRF prior, (iv) $\gamma(\bm{s})$ has a MGRF prior but uses PCA (see Section~\ref{sec:simulation_multivariate}),
(v) $\gamma(\bm{s})$ is orthogonal to the subspace spanned by the covariates, i.e., we use RSR. As we do not know which values are expected for $\rho$, we follow the recommendation in Section~\ref{sec:re_2} and use reformulation II for sampling the GRFs.
Inference is based on
100000 MCMC samples with burn-in of 50000 and thinning of 20.

\subsection{January 1, 2000}
Table~\ref{tab:res1} shows the posterior summaries for the covariates based on the data of January 1, 2020.
The results for the non-spatial model show that both elevation and minimum temperature have the expected effects and the equal-tailed $95\%$ credible intervals (CIs) do not include zero. Comparatively to the non-spatial model, in the base model the posterior mean of $\beta_1$ approximately doubles (from 0.369 to 0.792). However, the CI for minimum temperature includes zero.
By using our MGRF prior, the coefficient for $\beta_1$ decreases to 0.644 in comparison to the one of the base model. The minimum temperature has the expected negative effect and its CI does not cover zero.
This implies that, keeping everything else constant, a one standard deviation increase in elevation is associated with, on average, a 0.644 standard deviation increase in precipitation. A one standard deviation increase in minimum temperature is associated with, on average, a 0.137 decrease in precipitation. The results of RSR are quite close to the MGRF's and thus substantiate our results.
 The results for MGRF-PCA are similar to the base model's.
 
 The posterior mean of $\rho$ in MGRF prior is approximately 0.137 and significant. For MGRF-PCA it is 0.039 and insignificant. Moreover, for MGRF the $r_{\bm{z}^*} \approx 0.36$ and $r_{\bm{\gamma}} \approx 0.14$. For MGRF-PCA, $r_{\bm{z}_{PCA}} \approx 0.55$ and $r_{\bm{\gamma}} \approx 0.14$. According to the results from Section~\ref{sec:simulation}, this indicates that we are indeed in a scenario where spatial confounding might arise and our prior structure might reduce confounding bias. Given the consistency of results between non-spatial, RSR and MGRF spatial model, together with the fact that GMRF-PCA is so close to the results from the base model, we have reasons to believe we are in a scenario similar to Scenario 2 in Section~\ref{sec:simulation_multivariate}, with low sample size and $r_{\bm{z}_i} > r_{\bm{\gamma}}$ for $i=1$ and/or $i=2$.

\subsection{March 3, 2000}
Table~\ref{tab:res2} shows the posterior summaries for the covariates based on the data of March 3, 2000. All models are in agreeance and give the expected signs for the posterior mean of $\beta_1$ and $\beta_2$. Only the CI of the MGRF prior for minimum temperature includes zero.
Although positive, the posterior mean for elevation changes considerably between spatial and non-spatial models. Thus, a great part of the effect of elevation in the non-spatial model is attributed to the spatial random effect in spatial models. The posterior mean of the minimum temperature is similar for all models. 

The posterior mean of $\rho$ in MGRF prior is approximately -0.053 and insignificant. For MGRF-PCA it is 0.156 and insignificant. Moreover, for MGRF the $r_{\bm{z}^*} \approx 0.43$ and $r_{\bm{\gamma}} \approx 0.14$. For MGRF-PCA, $r_{\bm{z}_{PCA}} \approx 0.51$ and $r_{\bm{\gamma}} \approx 0.15$. This indicates that we are in a scenario where spatial confounding bias might arise. The fact that results are so consistent for all scenarios indicates there is low correlation between covariate information and the spatial random effect.

\subsection{Concluding remarks}
This example demonstrates the existence of spatial confounding in the sense that our prior structure noticeably changes the posterior distribution of $\beta_1$ and  $\beta_2$, compared to a spatial model with a base prior. However, there is also spatial confounding in the sense that adding a spatial effect to a non-spatial model changes the signal of $\beta_1$ \citep{hanks2015restricted,reich2006effects}.
Thus, it is important to be aware of spatial confounding when interpreting regression coefficients arising from a spatial model, otherwise, one might reach incorrect conclusions concerning the effect of the given covariates.
Future work that analyzes the effects of different marginal variances of the GRFs on the results in Section~\ref{sec:simulation}, should help us understand the need for standardizing variables.

\section{Discussion}\label{sec:discussion}
In spatial statistics, spatial confounding refers to the phenomenon where spatially varying covariates that model the mean of a response are correlated with the spatial random field that models spatial correlations.
We use the Bayesian framework to develop a prior structure against spatial confounding in continuously indexed spatial models. The MGRF prior exploits sparsity in two ways: (i) by using sparse precision matrices of GMRFs within the SPDE-approach, and (ii) by introducing a light parameterization in the case of multiple covariates that accounts for the correlation between a linear combination of the covariates -- instead of each covariate individually -- and the spatial random effect. The MGRF prior can be extended to different response distributions, by using the appropriate link function.

In a simulation study, we investigate the sources of spatial confounding, from the perspective of the spatial range. Future simulation studies should also address the effect of different marginal variances.
Second, we evaluate the properties of the proposed prior in a simulation study.
In the one covariate case, our prior structure outperforms a spatial model which does not account for spatial confounding, RSR and a non-spatial model. Compared to RSR, the MGRF prior additionally prevents overfitting, and provides the flexibility of user-defined scaling. In the multiple covariates case, the MGRF prior model either outperforms RSR or performs similarly well.
 On the negative side, in multiple covariates case, it also requires a case-by-case set-up.
Nonetheless, this extension is still strong enough to reveal the shortcomings of RSR.

 In an application, we consider daily precipitation in Germany and explore the effect of using the MGRF prior on the estimated regression coefficients for minimum daily temperature and elevation. This application gives evidence to the fact that regression coefficients can be biased if we do not account for spatial confounding.

Future research steps might follow two directions: (i) the model developed is used as a testing tool, or (ii) the model stays in its current format as standalone estimation method for spatial models.
In the former, the prior is used to test for the presence of spatial confounding. Then, one could use posterior evidence on the distance to the base model to decide on the presence of spatial confounding.

In future work, the multiple covariates case can benefit from more flexibility. While it is (computationally) convenient for the size $b^*$ of $\bm{\rho}$ to be smaller than the total amount of confounded covariates $B^*$, estimating only one correlation parameter might be too restrictive. Instead, if we ignore the correlation between the  covariates, we can consider $\bm{\rho} = (\rho_1, \ldots, \rho_{b^*})$ with $1 \leq b^* \leq B^*$. The value of $b^*$ and corresponding covariate transformation can be obtained in a pre-processing step that groups covariates using standard techniques from multivariate analysis such as PCA. This method can help separating negative and positive correlations.
 Nonetheless, as is, this option is not straightforward and it might be difficult to estimate multiple correlation parameters. In the case of the latter, one might investigate a hierarchical PC-prior structure that induces more shrinkage. Finally, one should reflect on how to address correlation between spatial covariates.

\bibliography{mybib}{}
\newpage

\begin{table}[ht!]
\centering
\medskip
\begin{tabular}{lcccc}
\toprule[0.09 em]
& \multicolumn{2}{c}{$\beta_1$ (elevation)} & \multicolumn{2}{c}{$\beta_2$ (temperature)}  \\
 \cmidrule{2-5}
Model & Mean & 95\% CI & Mean & 95\% CI\\
\midrule
Non-spatial & 0.369 & [0.323, 0.418] & -0.180  & [-0.230, -0.130]\\
Base spatial &  0.792  & [0.680, 0.904] & 0.003  & [-0.117, 0.113]\\
MGRF spatial & 0.644 &  [0.451,  0.810] & -0.137 & [-0.331, -0.030]\\
MGRF-PCA spatial & 0.835 & [0.702, 0.958] & 0.013 & [-0.111, 0.129] \\
RSR & 0.698 &  [0.610, 0.780] &  -0.098 & [-0.180, -0.022]\\
\bottomrule[0.09 em]
\end{tabular}
	\caption{Mean and equal-tailed 95\% credible interval (CI) for the posterior of $\beta_1$ and $\beta_2$ in the five models considered.}\label{tab:res1}
\end{table}

\newpage

\begin{table}[ht!]
\centering
\medskip
\begin{tabular}{lcccc}
\toprule[0.09 em]
& \multicolumn{2}{c}{$\beta_1$ (elevation)} & \multicolumn{2}{c}{$\beta_2$ (temperature)}  \\
Model & Mean & 95\% CI & Mean & 95\% CI\\
\midrule
Non-spatial & 0.368 & [0.291, 0.444] & -0.226 & [-0.341, -0.111]\\
Base spatial &  0.590 & [0.471, 0.704] & -0.226  & [-0.341, -0.111]\\
MGRF spatial & 0.666  &  [0.463,  0.856] & -0.142 & [-0.353, 0.053]\\
MGRF-PCA spatial & 0.625  &  [0.495,  0.749] & -0.282 & [-0.411, -0.156]\\
RSR & 0.594 &  [0.474,  0.706] &  -0.225 & [-0.340, -0.116]\\
\bottomrule[0.09 em]
\end{tabular}
	\caption{Mean and equal-tailed 95\% credible interval (CI) for the posterior of $\beta_1$ and $\beta_2$ in the five models considered.}\label{tab:res2}
\end{table}

\end{document}